\documentclass{article}

\usepackage{PRIMEarxiv}

\usepackage[utf8]{inputenc} 
\usepackage[T1]{fontenc}    
\usepackage{hyperref}       
\usepackage{url}            
\usepackage{booktabs}       
\usepackage{amsfonts}       
\usepackage{nicefrac}       
\usepackage{microtype}      
\usepackage{lipsum}
\usepackage{fancyhdr}       
\usepackage{graphicx}       
\graphicspath{{media/}}     

\usepackage{multirow}
\usepackage{float} 
\usepackage{tabularx}
\usepackage{caption}
\usepackage{amsmath}

\usepackage{subcaption}

\title{Scalable, Trustworthy Generative Model for Virtual Multi-Staining from H\&E Whole Slide Images}

\author{
  Mehdi Ounissi$^{1}$, Ilias Sarbout$^{1}$, $^{3}$Jean-Pierre Hugot, $^{4}$Christine Martinez-Vinson,\\ \textbf{Dominique Berrebi}$^{2,*}$ \textbf{and Daniel Racoceanu}$^{1,*}$\\
  $^{1}$Sorbonne Université, $^{1}$CNRS, $^{1,2}$Inserm, $^{1,2,3,4}$AP-HP, $^{1}$Inria, $^{1}$Paris Brain Institute-ICM\\
  $^{2,3}$Paris Cité University, $^{2}$Necker-Enfants Malades University Hospitals\\
  $^{2}$Anatomopathology Department, $^{2,3}$Centre de recherche sur l’inflammation\\
  $^{2,3,4}$Robert Debré University Hospitals\\
  Paris, France\\
  \small{$^*$Corresponding authors}
}

\begin{document}
\maketitle

\begin{abstract}

Chemical staining methods, while reliable, are time-consuming and can be resource-intensive, involving costly chemical reagents and raising environmental concerns. This underscores the compelling need for alternative solutions like virtual staining, which not only accelerates the diagnostic process but also enhances the flexibility of stain applications without the associated physical and chemical costs. Generative artificial intelligence technologies prove immensely useful in addressing these challenges. However, in healthcare, particularly within computational pathology, the high-stakes nature of decisions complicates the adoption of these tools due to their often opaque processes. Our work introduces an innovative approach that harnesses generative models for virtual stain transformations, enhancing performance, trustworthiness, scalability, and adaptability within computational pathology. The core of the proposed methodology involves a singular Hematoxylin and Eosin (H\&E) encoder that supports multiple stain decoders. This design prioritizes critical regions in the latent space of H\&E tissues, leading to a richer representation that enables precise synthetic stain generation by the decoders. Tested to simultaneously generate eight different stains from a single H\&E slide, our method also offers significant scalability benefits for routine use by loading only necessary model components during production. We integrate label-free knowledge during training, leveraging loss functions and regularization to minimize artifacts, thereby enhancing the accuracy of virtual staining in both paired and unpaired settings. To build trust in these synthetic stains, we employ a real-time self-inspection methodology using trained discriminators for each stain type, providing pathologists with confidence heat-maps to aid in their evaluations. Additionally, we perform automatic quality checks on new H\&E slides to ensure they conform to the trained H\&E distribution, guaranteeing the generation of high-quality synthetic stained slides. Recognizing the challenges pathologists face in adopting new technologies, we have encapsulated our method in an open-source, cloud-based proof-of-concept system. This system enables users to easily and virtually stain their H\&E slides through a browser, eliminating the need for specialized technical knowledge and addressing common hardware and software challenges. It also facilitates real-time user feedback integration. Lastly, we have curated a novel dataset comprising eight different paired H\&E/stains related to pediatric Crohn’s disease at diagnosis, providing 30 whole slide images (WSIs) for each stain set (total of 480 WSIs) to stimulate further research in computational pathology.

\end{abstract}

\keywords{Computational Pathology \and Virtual staining \and Trustworthy Artificial Intelligence \and Annotation-free knowledge-guided learning \and Cloud-based digital pathology \and H\&E/Immunohistochemical dataset}

\begin{figure}[htbp]
  \centering
  \includegraphics[width=\textwidth]{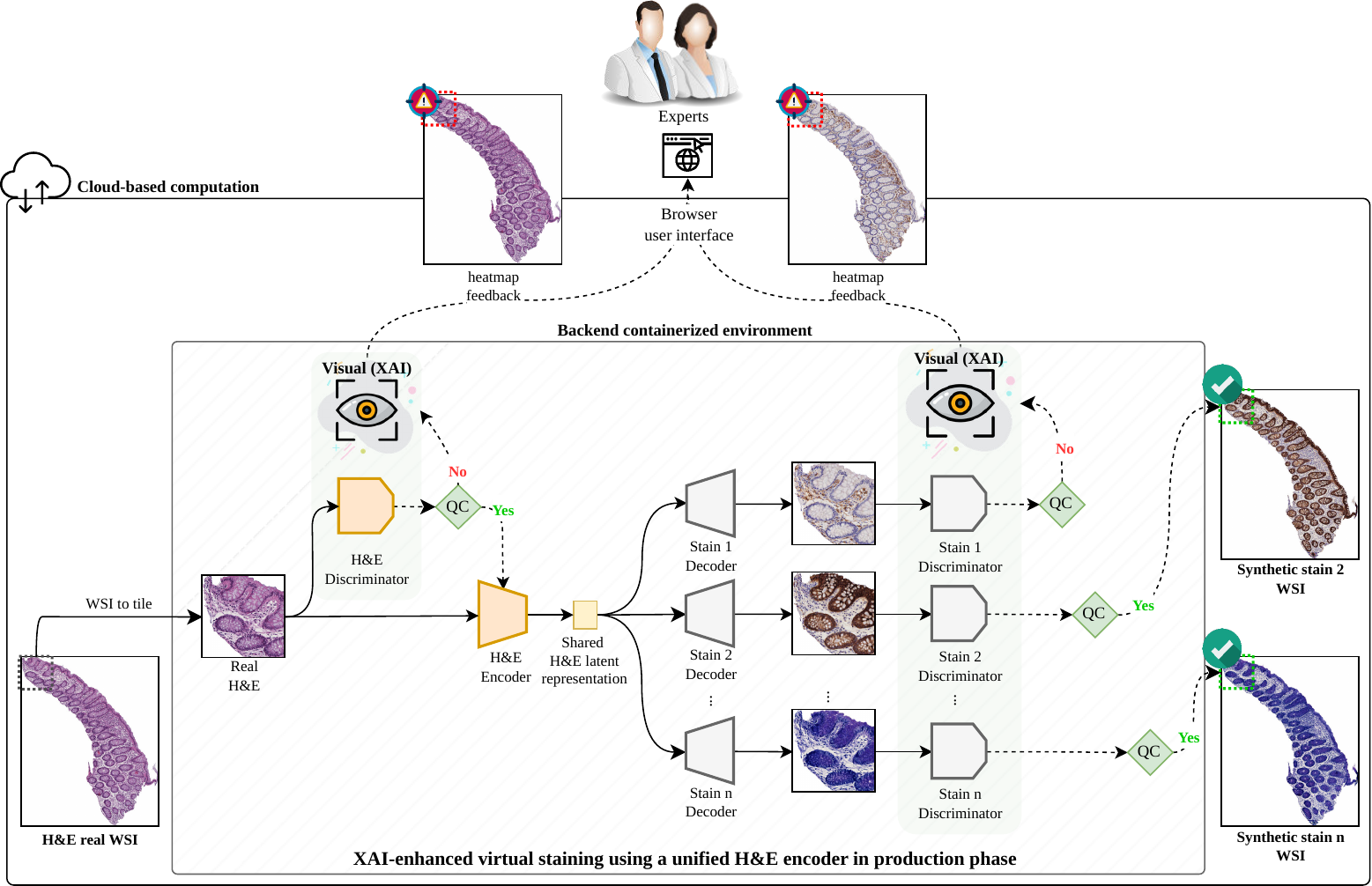}
    \caption{\textbf{Visual-XAI-enhanced trustworthy virtual staining approach.} End-to-end virtual staining approach generating synthetic IHC stains by using a single H\&E encoder and multiple stain decoders. Quality check (QC) protocol based on self-inspection features uses trained discriminators to consolidate trust in the generated synthetic stains, by ensuring the alignment of the new H\&E slides with the trained distribution and by validating the quality of the generated stained slides. Integration of cloud-based computing enhances accessibility and adoption by enabling pathologists to efficiently process large datasets from anywhere, while end-to-end system's algorithms are handled in a back-end containerized environment.}

  \label{fig:virtual-XAIinfrence}

\end{figure}

\newpage
\section*{Introduction}

Well-recognized standard practice in histopathology, Hematoxylin and Eosin (H\&E) staining offers multiple benefits, making it a traditionally preferred choice in histopathology, worldwide. Not only does this technique deliver efficient and cost-effective results, but it has also firmly established its place within reference routine diagnostic protocols in anatomopathology, reaching an indisputable central role in diagnostic protocols, such as the cancer grading~\cite{SAHA201829,Echle2021}. Thus, quantitative and qualitative measures on H\&E staining represent the reference for the design of treatment strategies. 

Despite these advantages, an intrinsic limitation of H\&E staining is represented by its constricted potential to identify specific proteins within cells - an endeavor often deemed crucial for pinpoint disease diagnosis and/or severity evaluation.

To compensate this drawback, Immunohistochemical (IHC) staining represents a widely accepted viable solution, by being particularly effective at identifying specific proteins within cells. This feature is crucial in the classification of various tumor types and is key in pinpointing the origin of metastatic tumors. Additionally, it can reveal minute tumor cells that might escape detection through standard staining procedures. This technique is particularly beneficial in diagnosing diseases that traditional biopsy cultures and serological diagnoses struggle to detect~\cite{Magaki2019,OumarouHama2022}.

Despite these strengths, IHC procedures have significant limitations. These methods require substantial resources, including time-intensive sample preparation and expert oversight, which increase the likelihood of errors and delays. Such delays could adversely affect disease diagnosis and treatment. Furthermore, the toxicity of the chemicals used in IHC can compromise further analysis of the same tissue samples and may pose environmental hazards~\cite{Bai2023}.

The inherent shortcomings of both H\&E and IHC staining techniques highlight the pressing need for an automated, digital, and reliable process, at least for the pre-selection of key stains from a wide array of possibilities. Such an enhanced process should aim to improve diagnostic precision while circumventing the associated time and monetary constraints.

In this context, the latest advances in computational pathology deserve special attention, particularly the utilization of deep learning methods. These methods can successfully convert H\&E stains into other IHC stains~\cite{Bai2023}. Additionally, deep learning techniques have demonstrated the capacity to generate synthetic IHC based on H\&E slides, a process that has been shown to enhance diagnostic accuracy~\cite{deHaan2021}.

Exhibiting potential, both supervised and unsupervised deep learning methods have demonstrated promising outcomes in transforming H\&E to IHC stains across diverse organ types~\cite{Borhani19,rivenson2019,Abraham22}. This transformation can be accomplished via two primary techniques: supervised translation, often termed as \textit{"paired"}, and unsupervised translation, referred to as \textit{"unpaired"}.

The paired translation method utilizes both H\&E and other chemically stained slides for transformation. This may involve using the same H\&E slide after washing and re-staining or using an adjacent slice. In contrast, the unpaired translation method does not require specific alignment between H\&E and stained slides. Together, these techniques demonstrate significant potential for advancing the field of computational pathology.

However, trust issues associated with existing methodologies, particularly deep generative models, remain a significant barrier. Clinicians and pathologists frequently find it difficult to rely on the predictions produced by these models, especially in real-world applications. Furthermore, the requirement for specialized hardware and software adds to the complexity. These technical requirements often pose challenges in seamlessly integrating computational techniques into routine pathology procedures.

We introduce a novel computational pathology pipeline that enhances the scalability, accuracy, trustworthiness, and utility of virtual staining techniques. Our approach includes a unified encoder serving multiple stain decoders, trust-building mechanisms via self-inspection, advanced training methodologies without additional annotations, cloud-based deployment, and a unique dataset focused on pediatric Crohn's disease. The distinguishing features of our research are structured around the next key contributions:

\begin{enumerate}
    \item \textbf{Unified H\&E encoder for multiple stain decoders:} We introduce a single, shared H\&E encoder that efficiently serves multiple decoders for generating diverse synthetic stains. This innovative architecture enhances learning capabilities and scalability by eliminating the need for multiple encoders. Extensive validation shows that our encoder can support up to eight distinct decoders simultaneously, significantly improving system performance and precision in synthetic stain generation.

    \item \textbf{Annotation-free knowledge guided training via loss functions and regularization:} Our methodology advances the training of H\&E encoders using specialized loss functions that require no additional annotations. These functions use existing stain data as a reference, penalizing inaccuracies more severely and enhancing model reliability. We have also tailored our approach for both paired and unpaired staining scenarios, ensuring accurate stain transformations across diverse conditions.

    \item \textbf{Trust enhancement in virtual stains via self-inspection and XAI heatmaps:} We prioritize reliability in synthetic stains with a dual mechanism of self-inspection and explainable AI (XAI). Our system uses discriminator models to assess the alignment of input slides with training data, as well as and heatmaps for real-time feedback, highlighting discrepancies in stain quality and fidelity. This approach empowers pathologists with tools for better decision-making and reinforces the trustworthiness of synthetic outputs.

    \item \textbf{Cloud-based virtual staining system:} We have successfully developed a proof-of-concept of our virtual staining method on an open-source, cloud-based platform. This innovative system allows pathologists to upload whole slide images (WSIs), generate synthetic stains, and provide feedback remotely. By eliminating the need for specialized hardware and software, this approach simplifies the use and accelerates the adoption of advanced staining technologies, providing guidelines that make them more accessible and user-friendly.

    \item \textbf{Novel H\&E/IHC paired stains dataset for pediatric Crohn's disease:} We curated a unique dataset of paired H\&E/special stains specific to pediatric Crohn's disease. This dataset, comprising 30 slides per stain type (480 slides in total), is designed to foster further research and development in the field.

\end{enumerate}

\newpage
\section{Related Works}

Virtual staining has been proposed to enable efficient transformations between different types of stains for WSIs. Over recent years, several improvements have been made to generate multiple stains using generative adversarial networks. However, substantial issues remain regarding scalability, accuracy, trustworthiness, and accessibility to clinicians~\cite{ciompi2017artificial, Bai2023}. This section explores the existing body of literature with an emphasis on the H\&E to IHC transformation process, aiming to highlight these current limitations and provide a comprehensive understanding of the landscape in which our research is situated.

\subsection{Stain synthesis using deep learning}
\label{sec:stain_SOA}

The field of computational pathology extensively researches stain transformations and synthesis. These studies aim to accurately digitally emulate tissue slide staining by using paired datasets, in which both H\&E stain and the corresponding WSI in other stains are included. There are several noteworthy studies in this field. \cite{deHaan2021} used a deep learning model that simultaneously processes H\&E tiles and outputs Jones, MT, and PAS stains. Similarly, \cite{Burlingame2020} developed the SHIFT method from a paired pancreas dataset to convert H\&E into virtual immunofluorescence images and estimate the tumor cell marker pan-cytokeratin distribution. Building on this, \cite{Hong2021} used a paired gastric carcinomas dataset to generate cytokeratin staining from H\&E to assist in diagnosing gastric cancer. \cite{Xie2022} applied a paired prostate dataset to transform H\&E to CK8 IHC stains, a preliminary step to reconstruct 3D segmented glands for prostate cancer risk stratification. \cite{liu2022bci} also developed a pyramid approach to generate human epidermal growth factor receptor 2 IHC stain from H\&E by using a paired breast cancer dataset. 

Despite these advancements, the paired H\&E/IHC staining method has some significant drawbacks. \cite{Yang2022} highlights that the staining processes are typically irreversible and present logistical and technical difficulties when acquiring paired data. In addition, inconsistencies in one type of staining can compromise the accuracy of the other, reducing the overall diagnostic value. 

To address these limitations, alternative approaches are needed. An example is C-DNN~\cite{Yang2022} method using cascaded deep neural networks to transform images from auto-fluorescence to H\&E, then to PAS, circumventing the challenge of acquiring paired data. Moreover, unpaired dataset settings have also been explored with the CycleGAN~\cite{goodfellow2014generative, cycleGAN}, a model widely employed. Unpaired datasets have been used in studies like \cite{levy2020preliminary} to transform H\&E to trichrome and \cite{mercan2020virtual} to convert H\&E and PHH3 stains. Other works include \cite{Lahiani2021perceptual}, which used a perceptual embedding consistency loss in GANs, and \cite{Liu2021unpaired}, which generated Ki-67-stained images from H\&E-stained images. Furthermore, the MVFStain framework~\cite{ZHANG2022MVFStain} was able to convert H\&E-stained images into multiple virtual functional stains in various scenarios.

In the literature, two principal methodological approaches for domain representation are distinguished. The first strategy employs separate pairs of encoders, decoders, and discriminators for each domain pairing, as exemplified by CycleGAN~\cite{cycleGAN} and its variants. This approach necessitates the training of \(3 \times n\) models, often resulting in scalability challenges during the training phase due to the extensive computational resources required. Conversely, methods such as StarGAN~\cite{stargan,stargan_2} utilize a unified model comprising a mapping network, a style encoder, a generator, and a discriminator. This configuration allows the generation of multiple latent domain representations, styled as distinct domains, which simplifies the training process by using a single model to accommodate multiple transformations.

However, these methodologies exhibit certain limitations, particularly in specialized applications. For instance, a pathologist requiring a subset of stains—specifically, \( s \) out of \( S \) potential stains created using H\&E—faces a substantial computational burden. They need to either load \( 2 \times s \) models (an encoder and a decoder for each required stain) or use an overarching model that encompasses all \( S \) stains. Both options demand considerable computational power, thereby delaying critical real-time responses. Furthermore, the current scope of virtual staining technology does not support the simultaneous synthesis of more than three IHC stains in a single session, a significant limitation when it comes to scaling and meeting the diverse needs of clinical environments.

Despite these challenges in digital pathology, the broader field of image processing has seen notable progress in addressing similar scalability issues. For instance, in the realm of unpaired art-style transfers, approaches like those presented in \cite{ComboGAN2017} employ a separate encoder, decoder, and discriminator for each domain, demonstrating substantial scalability potential. This success in other fields suggests that similar methodologies could be adapted for digital pathology, potentially enhancing scalability and efficiency in a domain where they are sorely needed.

The evolution of computational pathology has significantly benefited from diverse training strategies, loss functions, and regularization techniques. Works such as \cite{liu2018regularization,tellez2018whole} have contributed to substantial improvements in model performance. Nonetheless, embedding knowledge in a self-supervised manner without relying on additional labels continues to pose a complex challenge.

Moreover, the context within which computational pathology operates, particularly the selection of magnification in WSI interpretation, has gained increased attention. Studies like \cite{sirinukunwattana2016locality,courtney2018fully,KOSARAJU20203} have demonstrated the critical role of context in enhancing the performance of deep learning models for tissue characterization and cell classification. Yet, in the realm of virtual staining, there remains a significant gap with methods often relying on arbitrary magnification scale choices. This underscores the importance of further exploration in virtual staining techniques that utilize both paired and unpaired datasets, aiming to improve their applicability and effectiveness.

In conclusion, the profound influence of synthetic stains on patient outcomes necessitates these methods to be both efficient and trustworthy. Concerns related to their interoperability and consistency are pressing areas for improvement, which our work aims to address in the existing academic landscape.

\subsection{Virtual staining public datasets}\label{sec:data_sota}

In computational pathology, stain transformations and synthesis are pivotal research areas aimed at improving diagnostic accuracy. Many of these studies hinge on the use of diverse datasets. A large number of them employ paired datasets, which comprise both H\&E stains and their corresponding IHC stains on the same tissue slide. For instance, a notable study by~\cite{Burlingame2020} utilized a paired dataset of pancreas tissues to develop the SHIFT method, which transforms H\&E images into virtual PanCK immunofluorescence images, thereby estimating the tumor cell marker pan-cytokeratin's distribution. In a similar vein, \cite{deHaan2021} employed a dataset of paired tissue slides to transform H\&E tiles into Jones, Masson's Trichrome, and Periodic Acid–Schiff stains.

Furthermore, research has been extended to datasets featuring various types of cancers. \cite{Hong2021} used a paired dataset of gastric carcinomas to produce cytokeratin staining from H\&E, assisting in diagnosing gastric cancer. \cite{Xie2022} employed a paired prostate dataset for transforming H\&E to CK8 IHC stains, aiming to reconstruct 3D segmented glands for prostate cancer risk stratification. Meanwhile, \cite{liu2022bci} focused on generating the human epidermal growth factor receptor 2 (HER2) IHC stain from H\&E using a paired breast cancer dataset.

However, while paired datasets are instrumental, they are not without challenges. Since staining procedures are generally irreversible, acquiring such data can be technically challenging~\cite{Yang2022}. Recognizing these challenges, researchers have begun to explore unpaired datasets. For example, \cite{levy2020preliminary} successfully transformed H\&E to trichrome using an unpaired liver dataset and utilized a skin and lymph node dataset to change H\&E to SOX10 IHC. In another innovation, \cite{Lahiani2021perceptual} applied a perceptual embedding consistency loss in GANs, leveraging an unpaired liver dataset to morph H\&E into FAP-CK IHC stain. Beyond this, studies like \cite{Liu2021unpaired} have produced Ki-67-stained images from H\&E-stained ones using unpaired and unbalanced datasets from neuroendocrine tumors and breast cancers. The MVFStain~\cite{ZHANG2022MVFStain} framework is also notable, transforming H\&E-stained images into various virtual functional stains for tissues like mouse lung, breast cancer, and the rabbit cardiovascular system.

Yet, even with these strides in computational pathology, data availability remains a bottleneck, especially the scarce good quality public availability of paired H\&E/IHC stain datasets. To illustrate, although \cite{deHaan2021} publicly shared their approach's source code, the dataset they utilized remains private. Moreover, specific domains, such as pediatric Crohn's disease at the diagnosis stage (pre-treatment), remain under-researched and present opportunities for future exploration.

\subsection{Cloud-based computational pathology}\label{sec:cloud_sota}

Over recent years, the landscape of collaborative image analysis systems has witnessed significant advancements. A series of influential platforms emerged to reshape the domain. QuPath\cite{Bankhead2017} introduced the concept of web-based remote collaboration to computational pathology, allowing for annotation and the addition of modulable algorithms via Javascript and Groovy. Also, Open Reproducible Biomedical Image Toolkit (ORBIT)~\cite{Orbit} was launched, specializing in orchestrating existing analysis tools for medical imaging. Its collaborative capacities were enhanced by integrating OMERO\cite{OMERO_1,OMERO_2}. 

Despite limited AI capabilities in some tools, Cytomine\cite{cytomine} distinguishes itself with its innovative web-based interface. It was the first platform to enable the display of multiple WSIs in a web environment, eliminating the need for software installation. Cytomine's platform is notably comprehensive, incorporating all essential elements for server deployment—including web servers, job concurrency management, data storage, and a robust API. This integration makes it particularly suitable for histopathology applications.

Moreover, it enhances inclusivity and the reproducibility of results by supporting any dockerized algorithm. This feature allows authorized users access to a broad range of tools for collaborative medical image analysis. The platform's design facilitates job monitoring and enhances user interaction, which in turn improves collaboration and workflow management. Due to its effectiveness in promoting collaboration, managing medical image data efficiently, and integrating advanced machine learning techniques, it is increasingly favored for various applications.

To the best of our knowledge, no cloud-based open-source platform has previously incorporated virtual staining in a reliable manner. In response to this gap and in alignment with our technological advancements and research objectives, we have integrated our virtual staining method into the platform as a proof of concept. This integration provides a framework that empowers pathologists by eliminating the need for specific hardware and software requirements. It saves time and enhances their diagnostic and research capabilities in medical imaging analysis. This is achieved through a browser interface where all complex computations are managed in the backend, streamlining the user experience.
\section{Results}
\subsection{Study of unified vs. individual H\&E encoders in multi-virtual staining}

In our study, we aimed to develop an enhanced technique for generating multiple virtual stains simultaneously. Existing methods, as discussed in Section~\ref{sec:stain_SOA}, are limited to producing at most three stains concurrently. Inspired to improve upon these limitations, we utilized style transfer techniques from frameworks such as ComboGAN~\cite{ComboGAN2017}, which can handle up to 14 different art styles. We adapted this approach to histopathological applications, incorporating a novel architecture with a dedicated H\&E encoder, generator, and discriminator for the concurrent training of H\&E to multiple \(S\) stains, as illustrated in Section~\ref{sec:arch_DL} - Figs.~\ref{fig:virtual-XAIinfrence},~\ref{fig:train_unpaired_IHC},~and~\ref{fig:train_paired_IHC}.

Our results, detailed in Table~\ref{table:oneHE}, demonstrate the advantages of using a unified H\&E encoder for multi-virtual staining. Synthetic stains from this encoder consistently surpassed those from separate encoders for each stain. We evaluated performance using Mean Square Error (MSE), comparing the synthetic stains against authentic counterparts with a paired test set of H\&E samples. This comparison revealed that our method significantly outperforms the CycleGAN approach.

Both the unified and CycleGAN methods were tested under identical conditions, including the same dataset, training duration, and architecture for the encoders, generators, and discriminators. Our method not only offers significant gains in computational efficiency by employing a single encoder, decoder, and discriminator throughout the staining process but also requires fewer trainable parameters than the CycleGAN approach (refer to Figure~\ref{fig:unifed_vs_cyclegan_vs_stargan}). This streamlined architecture boosts computational efficiency and supports scalability, accommodating a broader range of output stains and accelerating the training process.

In conclusion, the unified H\&E encoder method excels by producing more accurate synthetic stains and achieving greater computational efficiency, making it a scalable and effective solution for large-scale histopathological studies.

\begin{table}[H]
    \centering
    \caption{\textbf{Performances and efficiency of the multi-virtual staining using unified H\&E encoder.} This table compares the Mean Squared Error (MSE) metrics (mean$\pm$std) of synthetic stain generation (unpaired setting) using our unified H\&E encoder versus traditional distinct H\&E encoders per stain (CycleGAN). The results highlight our method's superior accuracy and computational efficiency, featuring a significantly reduced number of trainable parameters, thus demonstrating its potential for scalable clinical-effective histopathological applications. For reproducibility details, refer to Section~\ref{exp:oneHE}.}
    
    \label{table:oneHE}
    \scriptsize
    \begin{tabularx}{\textwidth}{Xcccccccccc}
    \toprule
    \textbf{Method} & \textbf{AE1AE3} & \textbf{CD117} & \textbf{CD15} & \textbf{CD163} & \textbf{CD3} & \textbf{CD8} & \textbf{D240} & \textbf{GIEMSA} & \textbf{Overall MSE}$\downarrow$ & \textbf{Parameters}$\downarrow$ \\
    \midrule
    CycleGAN\cite{cycleGAN} & 0.1027 & 0.0504 & 0.0773 & 0.0660 & 0.0464 & 0.2659 & 0.0379 & 0.0321 & 0.0848$\pm$0.0717 & 409M\\
    \midrule
    Ours & 0.0854 & 0.0516 & 0.0674 & 0.0693 & 0.0463 & 0.0836 & 0.03 & 0.0365 & \textbf{0.0588$\pm$0.0209}& \textbf{230M}\\
    \bottomrule
    \end{tabularx}
\end{table}

\subsection{Annotation-free knowledge guided training and overall H\&E regularization}
\label{subsec:L_he+L_ihc}
To enhance the trustworthiness and robustness of virtual staining techniques in histopathology, we developed a novel methodology that incorporates additional constraints into the training model by leveraging information from stained slides. Contrary to style transfer applications in art~\cite{ComboGAN2017, stargan, stargan_2}, where domains differ significantly— simplifying the discriminator's task and placing greater emphasis on the generator for accurate image synthesis — the challenges in functional staining are rooted in the common morphological features across all stains, with variations primarily in activation reactions targeting specific proteins. This introduces two main challenges: (i) the model may underestimate regions with activated stains, which are statistically less frequent than non-activated tissue areas, leading to errors in stain generation where the discriminator struggles to differentiate between true activated regions and false negatives generated by the model; (ii) as the number of output stains increases, the encoder may disproportionately prioritize certain stains, potentially skewing the learning process and impeding overall performance.

\begin{figure}[H]
  \centering
  \includegraphics[width=\textwidth]{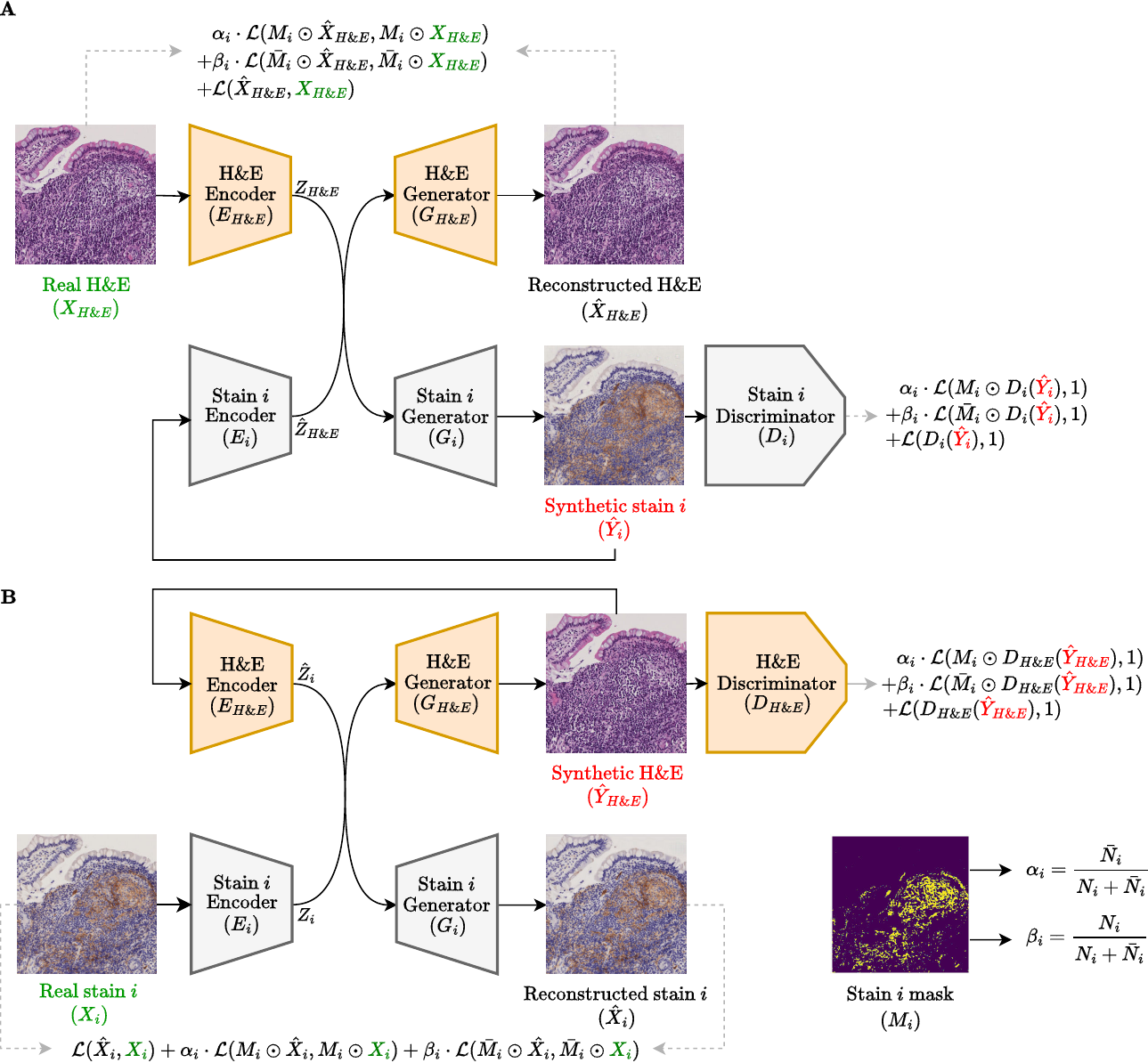}
  \caption{\textbf{Comprehensive representation of the training process for paired stain synthesis and computation of loss functions H\&E $\leftrightarrow$ stain \(i\).} \textbf{A.} Details the first training cycle, starting with a paired real H\&E image \(X_{H\&E}\) and generating a corresponding stain \(i\) image \(\hat{Y}_i\), followed by the reconstruction of the original H\&E image \(\hat{X}_{H\&E}\) to facilitate computation of the loss function components detailed in Section~\ref{sec:arch_DL}. \textbf{B.} Maps the second training cycle, beginning with a paired real stain \(i\) image \(X_i\), producing a corresponding H\&E image \(\hat{Y}_{H\&E}\), and concluding with the reconstructed stain \(i\) image \(\hat{X}_i\), using the staining mask \(M_i\) (\(\bar{M}_i\) corresponds to the complementary mask of \(M_i\)) to compute various elements of the loss function detailed in Section~\ref{sec:arch_DL}. Each panel illustrates the model’s modifications aimed at enhancing the precision and consistency of stain synthesis and discrimination in paired training scenarios.}
  \label{fig:train_paired_IHC}
\end{figure}

\begin{figure}[H]
    \centering
    \includegraphics[width=\textwidth]{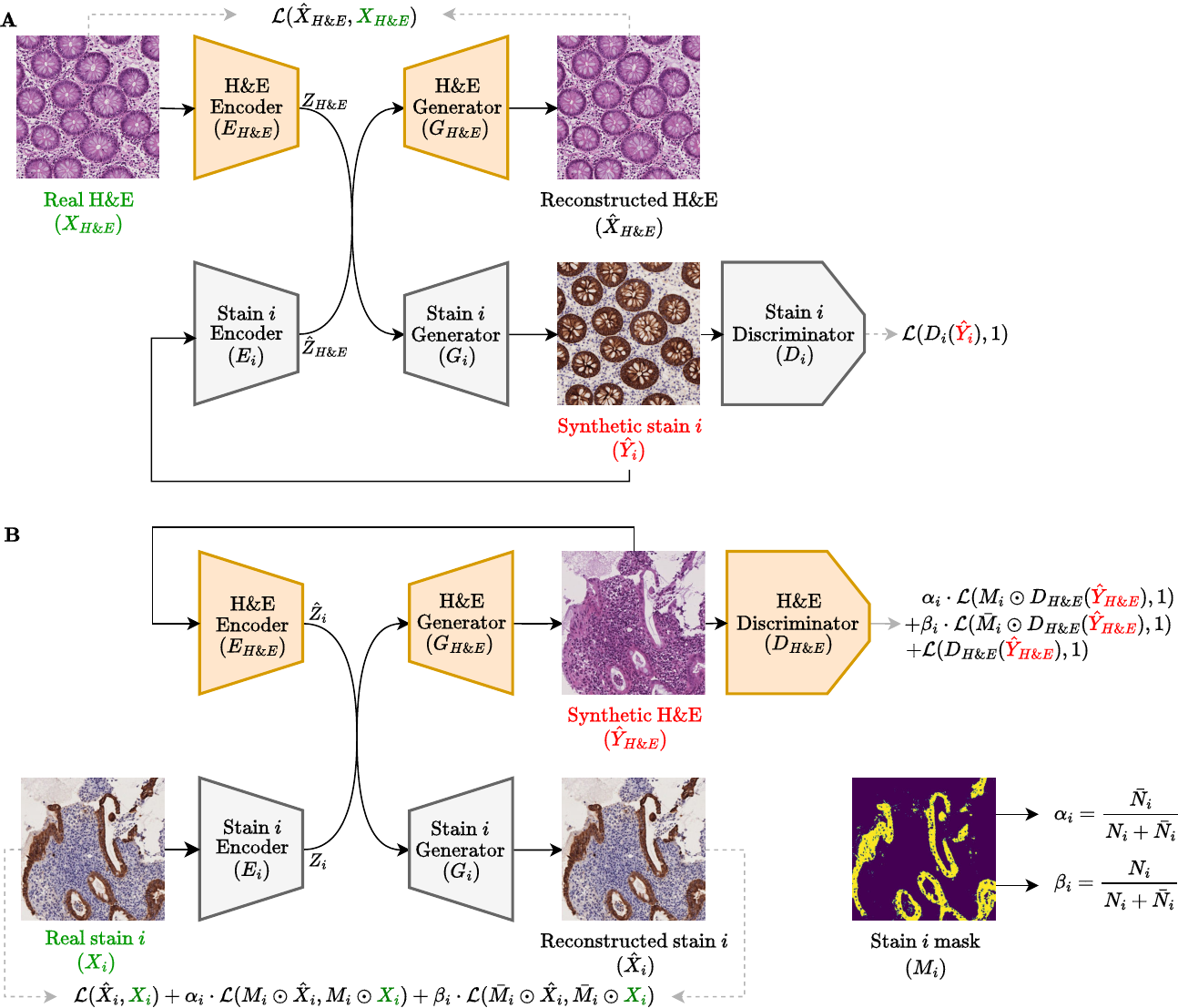}
    \caption{\textbf{Detailed representation of the scalable training process for unpaired stain synthesis and computation of loss functions H\&E $\leftrightarrow$ stain \(i\).} \textbf{A.} Illustrates the first training cycle, beginning with a real H\&E image \(X_{H\&E}\), generating a synthetic stain \(i\) image \(\hat{Y}_i\), and closing with the reconstructed H\&E image \(\hat{X}_{H\&E}\) to enable computation of the loss function components. \textbf{B.} Demonstrates the second training cycle, starting with a real stain \(i\) image \(X_i\), producing a synthetic H\&E image \(\hat{Y}_{H\&E}\), and concluding with the reconstructed stain \(i\) image \(\hat{X}_i\), incorporating the staining mask \(M_i\) (\(\bar{M}_i\) corresponds to the complementary mask of \(M_i\)) to compute various elements of the loss function (refer to Section~\ref{sec:arch_DL}). Each panel highlights different aspects of the model’s adaptations and refinements, targeting and enhancing underrepresented activated regions to ensure more accurate and consistent stain synthesis and discrimination.}
    \label{fig:train_unpaired_IHC}
\end{figure}

To address these challenges, our approach incorporates \(\mathcal{L}_{\text{IHC}}\) loss functions that automatically recognize stain-specific properties (see Figure.~\ref{fig:ihc_masks}), and adaptively modulate the loss functions to emphasize underrepresented activated regions (refer to Section~\ref{sec:arch_DL}, Figure.~\ref{fig:train_unpaired_IHC} and Figure.~\ref{fig:train_paired_IHC}), thereby minimizing errors and reducing hallucinations. Stain-activated regions are first identified and then used to spatially weight the loss functions, enhancing focus on relevant tissue areas.

Furthermore, we introduce an H\&E regularization \(\mathcal{L}_{\text{H\&E}}\) to maintain balanced attention across various stains, recalibrating the model to equally consider all stains by retro-propagating the mean error of generated stains through the H\&E components exclusively.

This strategy not only stabilizes the training process but also scales effectively, as demonstrated by the improved results in Table.\ref{table:immunoLoss}.The combination of \(\mathcal{L}_{\text{IHC}}\) with \(\mathcal{L}_{\text{H\&E}}\) enhances the model's performance in both paired and unpaired settings, ensuring consistent focus across different stains and significantly boosting overall efficacy.

\begin{table}
    \centering
    \caption{\textbf{Impact of incorporating \(\mathcal{L}_{\text{IHC}}\) loss Functions and \(\mathcal{L}_{\text{H\&E}}\) regularization on stain synthesis quality.} Comparative results displayed for paired and unpaired staining settings, quantified by MSE, peak signal-to-noise ratio (PSNR), and structural similarity index (SSIM). For reproducibility details, refer to Section~\ref{exp:immunoLoss}.}

    \label{table:immunoLoss}
    \begin{tabular}{cccccc}
    \toprule
      & \multicolumn{2}{c}{Synthesis loss functions}                           & \multicolumn{3}{c}{Metrics} \\
    \midrule
    Setting  & \(\mathcal{L}_{\text{IHC}}\) & \(\mathcal{L}_{\text{H\&E}}\) & MSE($\times10^{-2}$)$\downarrow$ & PSNR$\uparrow$   & SSIM ($\%$)$\uparrow$     \\
    \midrule
    Paired   &                              &                               & 2.413$\pm$0.951 & 21.22$\pm$2.51 & 83.60$\pm$7.7 \\
             &                              & \checkmark     & 1.768$\pm$0.572 & 22.51$\pm$2.49 & 85.10$\pm$7.1 \\
             & \checkmark    &                               & 1.673$\pm$0.579 & 22.78$\pm$2.51 & 86.38$\pm$6.8 \\
             & \checkmark    & \checkmark     & \textbf{1.658$\pm$0.586} & \textbf{22.85$\pm$2.60} & \textbf{86.37$\pm$6.9} \\
    \midrule
    \midrule
    Unpaired &                              &                               & 2.451$\pm$0.903 & 21.10$\pm$2.31 & 83.59$\pm$7.5 \\
             &                              & \checkmark     & 2.921$\pm$0.748  & 20.21$\pm$1.62 & 79.50$\pm$7.8 \\
             & \checkmark    &                               & 2.413$\pm$0.892 & 21.17$\pm$2.31 & 83.68$\pm$7.6 \\
             & \checkmark    & \checkmark     & \textbf{2.329$\pm$0.871} & \textbf{21.34$\pm$2.39} & \textbf{83.79$\pm$7.6} \\
    \bottomrule
    \end{tabular}
\end{table}

\begin{figure}
    \centering
    \includegraphics[width=1\linewidth]{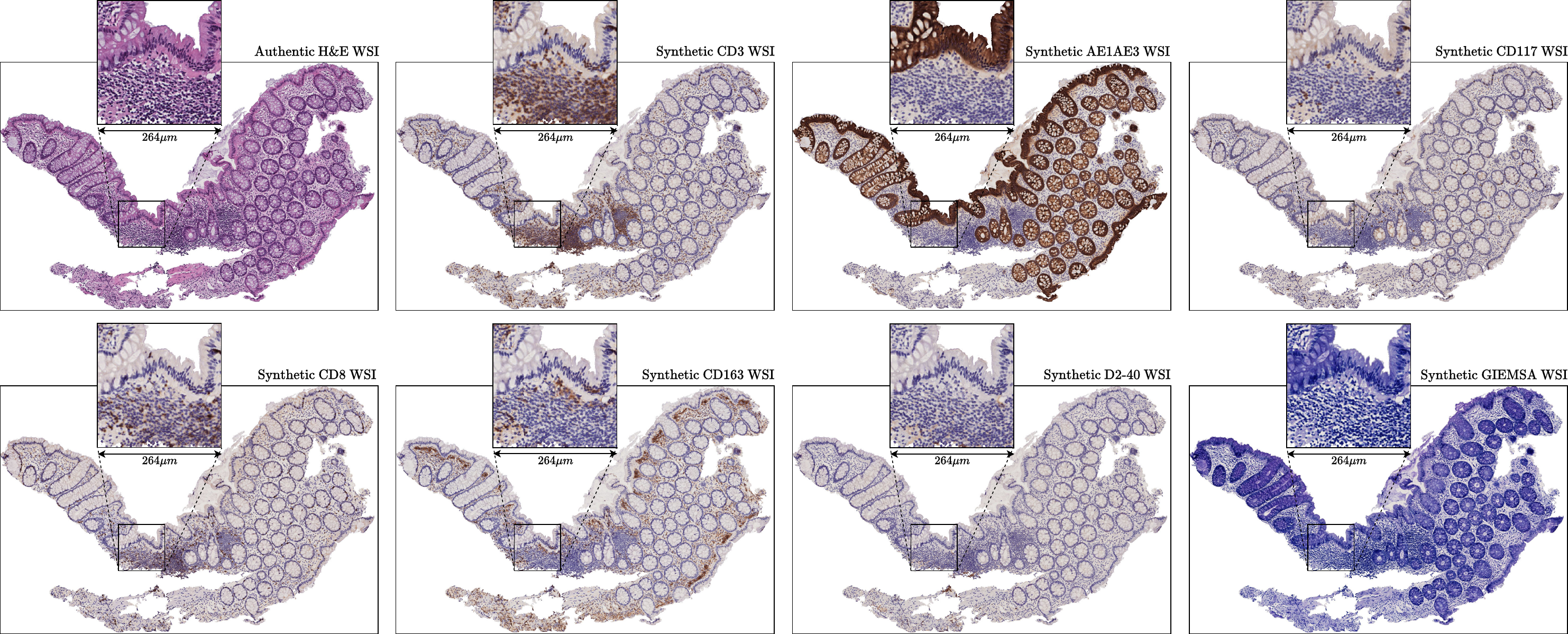}
    \caption{\textbf{Multi-virtual staining results in the context of Crohn's disease.} This figure showcases the high-resolution WSIs of various synthetic stains achieved using the \(\mathcal{L}_{\text{IHC}}\) and \(\mathcal{L}_{\text{H\&E}}\) loss functions in the unpaired setting.}
    \label{fig:demo_crohn}
\end{figure}

\subsection{Context-importance in multi-virtual staining quality and scalability}\label{results:context_scalability}
Due to the enormous size of WSIs, which typically measure around 10,000x10,000 pixels, current GPUs cannot process an entire slide at once during training. As a result, virtual staining techniques often utilize a sliding window tiling approach. This method involves dividing the slide into smaller, more manageable patches that are compatible with deep learning models and GPU capacities. Common dimensions for these patches are 128x128 pixels and 256x256 pixels~\cite{ZHANG2022MVFStain, UMDST}. Employing this approach necessitates careful consideration of the optimal magnification level (x10, x20, and x40) for analysis. The choice of magnification impacts the training process in paired and unpaired learning scenarios. Using smaller patches increases the total number of patches per WSI, raising questions about inference time -- specifically, the duration required to reconstruct a virtually stained slide. Addressing these technical issues is crucial not only for optimizing the performances but also for understanding the practical implications related to the inference time, a critical factor for pathologists.

In our empirical experiments, by leveraging a modular approach that does not require loading all model parts simultaneously, we were able to process 512x512 tiles while simultaneously outputting 8 stains plus H\&E during training on a standard 16GB GPU. This setup provides at least four times more spatial resolution than those reported in~\cite{ZHANG2022MVFStain, UMDST}, thereby offering more flexibility and the ability to incorporate more context within each patch. To determine the optimal magnification level for virtual staining, we trained our model using various magnifications, each resized to a uniform dimension of 512x512 pixels for consistent image processing. The specific magnifications tested were x10 (original tile size of 2048x2048 pixels $\approx$ 450.56x450.56$\mu m$), x20 (original tile size of 1024x1024 pixels $\approx$ 225.28x225.28$\mu m$), and x40 (original tile size of 512x512 pixels $\approx$ 112.64x112.64$\mu m$). These experiments were conducted under both paired and unpaired learning settings to evaluate the impact of magnification on model performance. As shown in Table~\ref{table:mag}, in the paired setting, all magnifications yielded comparable results due to the direct correspondence between the H\&E-stained WSI and other stained WSIs. Our experiments in unpaired settings revealed a valuable insight: lower magnifications, which provide broader contextual views, enhance the performances. This suggests that embracing more extensive contextual information is crucial for effective learning -- where direct stain correspondences are lacking --, guiding future improvements in virtual staining techniques.

\begin{table}
\centering
\caption{\textbf{Analysis of our model's performance across different magnifications.} This table presents the performance of our modular training approach at different magnifications (x10, x20, x40), where tiles extracted at each magnification were resized to 512x512 pixels to ensure a consistent image size for analysis. The models were trained using \(\mathcal{L}_{\text{IHC}}\) and \(\mathcal{L}_{\text{H\&E}}\) loss functions. In the paired setting, no particular magnification preference was observed, indicating uniformity in performance. Conversely, in the unpaired setting, lower magnifications, which provide more contextual information, demonstrated a significant advantage, underscoring the importance of context for effective learning in scenarios lacking direct correspondence between stain types. For reproducibility details, refer to Section~\ref{exp:mag}.}\label{table:mag}
\begin{tabular}{ccccccccccc}
\toprule
                           & \multicolumn{3}{c}{Trained on} & \multicolumn{3}{c}{Tested on} & \multicolumn{4}{c}{Metrics} \\
\midrule
Setting   & x10     & x20    & x40 & x10     & x20    & x40 & MSE($\times10^{-2}$)$\downarrow$ & PSNR$\uparrow$ & SSIM$\uparrow$ & Inference time$\downarrow$ \\
\midrule
Paired & \checkmark & & & \checkmark  & &  & \textbf{1.658$\pm$0.586} & 22.85$\pm$2.60 & 86.37$\pm$6.9 & \textbf{14.06 sec} \\
        &  & \checkmark &  &  & \checkmark & & 1.681$\pm$0.618 & 22.84$\pm$2.79 & 85.42$\pm$ 6.8 & 19.81 sec\\
        & & & \checkmark &   & &  \checkmark & 1.714$\pm$0.636 & 22.75$\pm$2.80 & \textbf{87.22$\pm$5.3} & 49.67 sec\\
     
     \cmidrule{2-11}

       &  & \checkmark &  & \checkmark & &  & 1.659$\pm$0.632 & \textbf{22.94$\pm$2.89} & 86.46$\pm$7.1 & 19.81 sec\\
       & & & \checkmark & \checkmark  & &   & 1.673$\pm$ 0.640 & 22.90$\pm$2.90 & 86.50$\pm$7.1 & 49.67 sec\\

\midrule
\midrule
Unpaired & \checkmark & & & \checkmark  & &  & \textbf{2.329$\pm$0.871} & \textbf{21.34$\pm$2.39} & \textbf{83.79$\pm$7.6} & \textbf{14.06 sec} \\
        &  & \checkmark &  &  & \checkmark &   & 3.476$\pm$2.227& 20.06$\pm$3.25 & 79.60$\pm$10.8  & 19.81 sec\\
        & & & \checkmark &   & &  \checkmark &  4.498$\pm$3.668 & 19.54$\pm$4.20 & 80.59$\pm$ 9.7 & 49.67 sec\\

         \cmidrule{2-11}
         &  & \checkmark &  & \checkmark & &   & 4.389$\pm$3.639 & 19.72$\pm$4.30 & 76.45$\pm$15.7 & 19.81 sec\\
        & & & \checkmark & \checkmark  & &   & 3.639$\pm$2.226 & 20.22$\pm$3.33 & 79.64$\pm$12.8 & 49.67 sec\\
\bottomrule
\end{tabular}
\end{table}

In the paired analysis, we initially utilized a 10x magnification, corresponding to a resolution of 512 x 512 pixels (approximately 0.88 µm per pixel). We further processed the original images by resizing them to 1024 x 1024 pixels (0.44 µm per pixel) to better evaluate the impact of pixel density on high-context paired training. To leverage the full capabilities of high-end GPUs, such as the NVIDIA A100 80GB, we also experimented with a maximum image size of 1400 x 1400 pixels (0.32 µm per pixel) during the training phase.

\begin{table}
\centering
\caption{\textbf{Scalability of our multi-virtual staining approach at various training resolutions in a paired setting.} This table displays the outcomes of training our virtual staining model on images with eight stains plus H\&E at different resolutions. The results demonstrate consistent performance across various pixel densities. The data highlights our approach's effective use of advanced GPU resources, emphasizing the scalability of our methodology.}

\label{table:res}
\begin{tabular}{cccccccccc}
\toprule
\multicolumn{3}{c}{Trained on patch size} & \multicolumn{3}{c}{Tested on patch size} & \multicolumn{4}{c}{Metrics} \\
\midrule
$512^2$     & $1024^2$    & $1400^2$ & $512^2$     & $1024^2$    & $1400^2$ & MSE($\times10^{-2}$)$\downarrow$    & PSNR$\uparrow$ & SSIM$\uparrow$ & Inference time$\downarrow$ \\
\midrule
\checkmark & & & \checkmark  & &  & \textbf{1.745$\pm$0.626} &	\underline{22.63$\pm$2.61} &	\underline{86.24$\pm$6.8} & \textbf{14.06 sec} \\
& \checkmark &  &  & \checkmark & & 1.766$\pm$0.654 &	22.61$\pm$2.74 &	85.27$\pm$6.8 & \underline{22.04 sec}\\
& & \checkmark &   & &  \checkmark & 1.87$\pm$0.723 &	22.40$\pm$2.79	& 85.43$\pm$6.6  & 37.51 sec\\
     
\midrule

& \checkmark &  & \checkmark & &  & \underline{1.748$\pm$0.667} & \textbf{22.70$\pm$2.84} & \textbf{86.33$\pm$7.0} & \underline{22.04 sec}\\
& & \checkmark & \checkmark  & &   & 1.853$\pm$0.714 & 22.45$\pm$2.84 & 86.11$\pm$7.1 & 37.51 sec\\

\bottomrule
\end{tabular}
\end{table}

Table.\ref{table:res} illustrates the scalability of our modular approach, which can successfully process images with eight stains plus H\&E up to the 1400 x 1400 resolution. The comparative results in Tables \ref{table:mag} and \ref{table:res} emphasize that enhancing the contextual information within the images proves substantially more advantageous than increasing pixel density.

\subsection{Regularization techniques impact on unpaired multi-virtual staining quality}

Most style transfer methods emphasize the importance of regularization to improve and stabilize the training process. For example, identity mapping loss regularization is critical in preserving the color of input paintings in artistic applications, as seen in the CycleGAN~\cite{cycleGAN} framework. Similarly, forward loss regularization plays a crucial role in virtual staining by preserving morphological features when translating from H\&E staining to other types, as highlighted in the UMDST~\cite{UMDST} model. Despite the variety of available regularization techniques, comprehensive ablation studies to evaluate their effectiveness in virtual staining are lacking.

In our research, presented in Table.\ref{table:compReg}, we conducted an extensive ablation study to assess the individual and combined effects of various regularization techniques on the quality of virtual stain synthesis. This study examines different combinations of synthesis loss functions and regularization methods to identify the most effective configurations. The metrics used in the study include MSE, PSNR, and SSIM, which gauge the error, quality, and visual similarity of the synthesized images, respectively.

Our results, as presented in Table.\ref{table:compReg}, offer a detailed analysis of how various combinations of loss functions—specifically identity loss \(\mathcal{L}_{\text{idt}}\), latent loss \(\mathcal{L}_{\text{lat}}\), and forward loss \(\mathcal{L}_{\text{fwd}}\) (refer to Section~\ref{method:stain_regs})—impact key performance metrics such as MSE, PSNR, and SSIM. Each row in the table represents a different combination of these loss functions, illustrating their respective effects on the evaluation metrics. This thorough evaluation provides essential insights into the efficacy of each approach.

Firstly, applying the forward loss \(\mathcal{L}_{\text{fwd}}\) alone has demonstrated superior results compared to baselines that either include or exclude the combination of \(\mathcal{L}_{\text{IHC}}\) and \(\mathcal{L}_{\text{H\&E}}\). The \(\mathcal{L}_{\text{fwd}}\) is particularly effective as it assists the model in preserving the morphological features that are highlighted by the H\&E staining, crucial for accurate virtual staining.

Secondly, the best performance is achieved when \(\mathcal{L}_{\text{fwd}}\) is combined with \(\mathcal{L}_{\text{idt}}\). This combination not only preserves the morphological integrity of the stains but also maintains the original features of the input images, thereby ensuring a high fidelity in the virtual staining process. This suggests that integrating both forward and identity losses provides a robust method for enhancing the quality and accuracy of the synthesized stains, making it particularly suitable for applications requiring high precision in unpaired virtual staining.

Conversely, the inclusion of latent loss \(\mathcal{L}_{\text{lat}}\) in the combinations tested does not appear to contribute positively to the staining outcomes. In fact, configurations incorporating \(\mathcal{L}_{\text{lat}}\) consistently under-performed in all metrics compared to those without it. This observation suggests that latent loss may interfere with the preservation of crucial staining characteristics -- specific to each different stain --, thus making it a less desirable option in the context of virtual staining where accuracy and fidelity are paramount.

\begin{table}
\centering
\caption{\textbf{Effects of various regularization techniques on unpaired virtual staining performance.} This table displays an ablation study of different combinations of synthesis loss functions (\(\mathcal{L}_{\text{IHC}}\), \(\mathcal{L}_{\text{H\&E}}\) detailed in Sections~\ref{method:Lcyc_Lihc}~and~\ref{method:Lhe}) and regularization methods (\(\mathcal{L}_{\text{idt}}\), \(\mathcal{L}_{\text{lat}}\)and \(\mathcal{L}_{\text{fwd}}\) detailed in Section~\ref{method:stain_regs}) on the performance metrics MSE, PSNR, and SSIM. Each row represents a specific configuration of loss functions, illustrating their impact on the accuracy and quality of virtual staining outcomes. For reproducibility details, refer to Section~\ref{exp:compReg}.}
\label{table:compReg}
\begin{tabular}{cccccccc}
\toprule
\multicolumn{2}{c}{Synthesis loss functions}  & \multicolumn{3}{c}{Regularization}                         & \multicolumn{3}{c}{Metrics} \\
\midrule
\(\mathcal{L}_{\text{IHC}}\) & \(\mathcal{L}_{\text{H\&E}}\) & \(\mathcal{L}_{\text{idt}}\) & \(\mathcal{L}_{\text{lat}}\) & \(\mathcal{L}_{\text{fwd}}\) & MSE($\times10^{-2}$)$\downarrow$    & PSNR$\uparrow$ & SSIM ($\%$)$\uparrow$   \\
\midrule
              &             &  & & & 2.451$\pm$0.903 & 21.10$\pm$2.31 & 83.59$\pm$7.5 \\
\midrule
\checkmark    & \checkmark  &    & & & 2.329$\pm$0.871 & 21.34$\pm$2.39 & 83.79$\pm$7.6 \\

\midrule
\checkmark    & \checkmark  & \checkmark    & & & 2.378$\pm$0.850 & 21.23$\pm$2.32 & 83.93$\pm$7.5 \\
\checkmark    & \checkmark  &    &\checkmark  & & 2.477$\pm$0.940 & 21.08$\pm$2.38 & 83.32$\pm$7.7 \\
\checkmark    & \checkmark  &    &  & \checkmark & \underline{2.258$\pm$0.845} & \underline{21.50$\pm$2.50} & \underline{84.28$\pm$7.6} \\
\midrule
\checkmark    & \checkmark  &  \checkmark   & \checkmark  & & 2.459$\pm$0.952 & 21.13$\pm$2.46 & 83.43$\pm$7.6\\
\checkmark    & \checkmark  &  \checkmark   &   & \checkmark & \textbf{2.244$\pm$0.797} & \textbf{21.51$\pm$2.45} & \textbf{84.31$\pm$7.6} \\
\checkmark    & \checkmark  &    &  \checkmark  & \checkmark & 2.315$\pm$0.897 & 21.41$\pm$2.51 & 84.19$\pm$7.5 \\
\midrule
\checkmark    & \checkmark  &  \checkmark   &  \checkmark  & \checkmark & 2.488$\pm$0.882 & 21.03$\pm$2.23 & 83.23$\pm$7.6 \\
\bottomrule
\end{tabular}
\end{table}

\subsection{Mitigating stitching artifacts in WSI virtual staining}

\begin{figure}[H]
    \centering
    \begin{subfigure}[b]{\linewidth}
        \centering
        \includegraphics[width=0.8\linewidth]{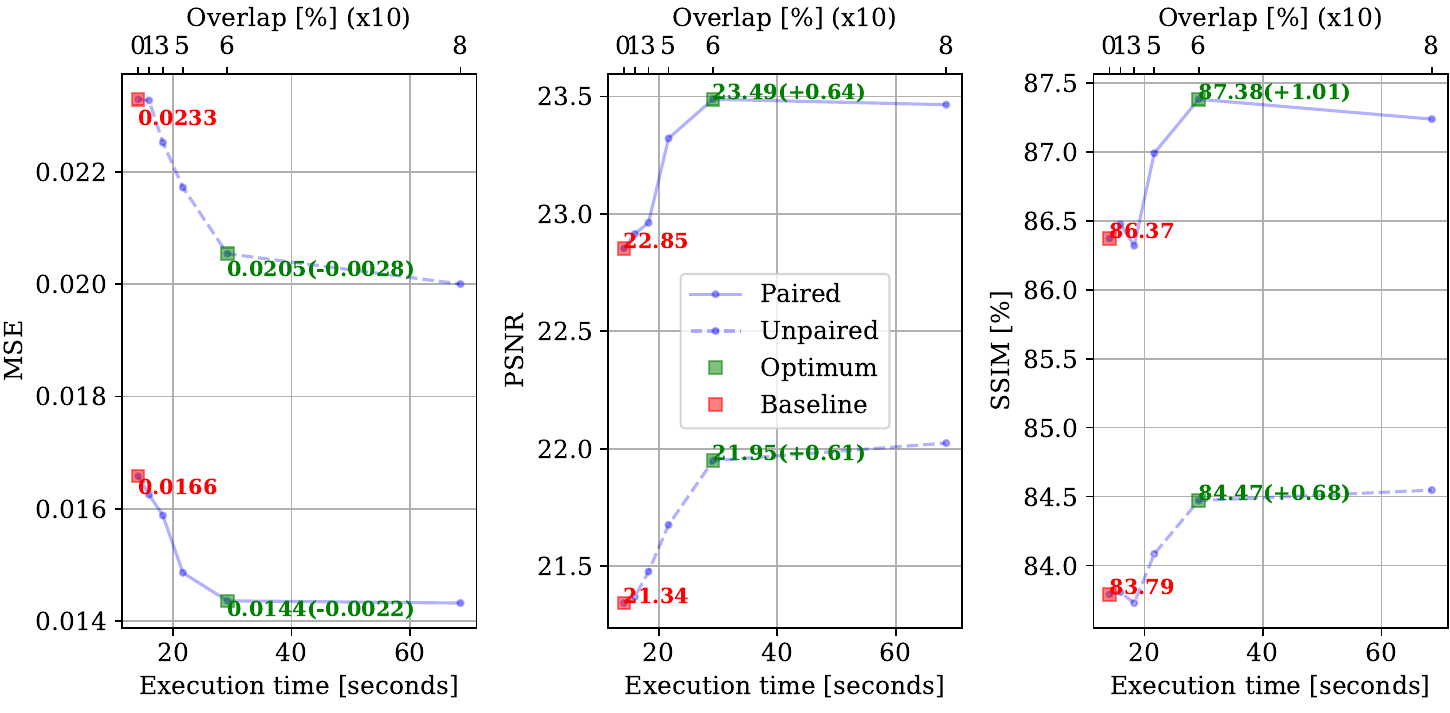}
        \caption{Overlap post-processing quantitative performance evaluation}
        \label{fig:overlap_performance}
    \end{subfigure}

    \vspace{1em} 

    \begin{subfigure}[b]{\linewidth}
        \centering
        \includegraphics[width=0.8\linewidth]{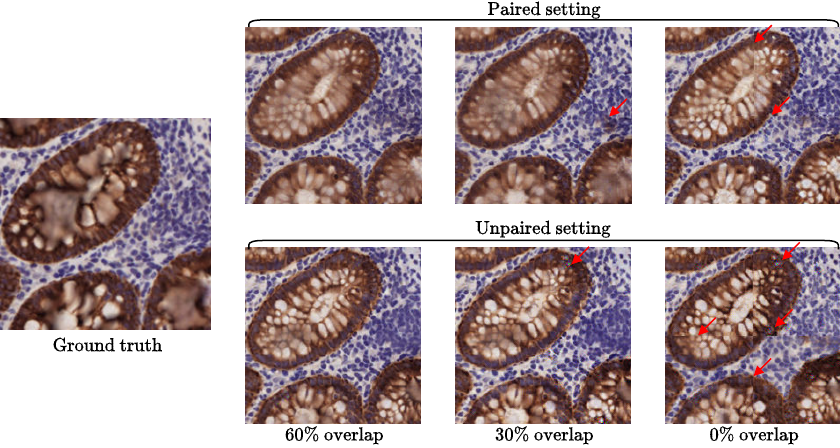}
        \caption{Overlap post-processing qualitative performance evaluation}
        \label{fig:overlap_effect}
    \end{subfigure}

    \caption{\textbf{Illustration of post-processing effects on stitching artifacts and performance metrics in virtual staining.} \textbf{(a)} Depicts the improved outcomes using different overlap approaches with a Hamming window, emphasizing enhanced image quality and reduced artifacts, with the optimal performance-time execution ratio achieved at 60\% overlap. \textbf{(b)}~Shows typical stitching artifacts at tile borders with 0\%, 30\%, and 60\% overlaps, marked by red arrows, demonstrating the sudden color changes and errors near the boundaries. This figure highlights the comparison across performance metrics (MSE, PSNR, SSIM) in both paired and unpaired settings, showcasing the effectiveness of the post-processing strategy in enhancing the overall quality and facilitating the adoption of virtual staining technologies in clinical environments. For reproducibility details, refer to Section~\ref{method:tile_stitch}}
    \label{fig:overlap_performance_and_effect}
\end{figure}

In our review of existing virtual staining approaches~\cite{mercan2020virtual, Burlingame2020, Hong2021, Liu2021unpaired, deHaan2021,ZHANG2022MVFStain, UMDST}, both paired and unpaired, it is evident that most employ a sliding window tiling approach during model training, as discussed in Sections~\ref{sec:stain_SOA}~and~\ref{results:context_scalability}. This training method inevitably leads to challenges in reconstructing WSIs from the resultant patches. Particularly, it can cause visible artifacts such as sudden color changes at tile borders and errors near these boundaries, as demonstrated in Figure.\ref{fig:overlap_performance_and_effect}.b (0\% overlap in both settings, marked with red arrows). These artifacts not only undermine the trust of pathologists in the tools but can also increase cognitive load and error rates during slide examination. This issue is pervasive across all tile-based virtual staining methods, underscoring the necessity for a universal solution.

To mitigate these problems, we developed a post-processing technique tailored for WSIs in the context of virtual staining. Our observations indicate that the models are context-sensitive, performing with high accuracy at the tile's center and less so near the edges. Leveraging this insight, our approach involves stitching tiles with an intentional overlap that prioritizes the central regions of the tiles using a Hamming\cite{hamming1, hamming2} window (refer to Section~\ref{method:tile_stitch}), effectively enhancing performance without additional training. This method, depicted in Figure.\ref{fig:overlap_performance_and_effect}, significantly improves all evaluated metrics in both paired and unpaired settings and results in higher perceived image quality relative to the ground truth. To further refine our approach, it's important to note that while our post-processing technique does introduce a slight increase in processing time, it offers a significant benefit in terms of performance-to-time ratio. An optimal overlap of 60\%, as illustrated in the figures, provides the best balance between performance enhancement and execution time (>1min per 8 stains).This post-hoc processing strategy not only addresses the stitching artifacts effectively but also enhances the overall utility of virtual staining technologies in clinical settings. By maintaining a streamlined workflow, it facilitates the routine use of these technologies in fast-paced clinical environment, potentially broadening their adoption and building trust among pathologists. This adjustment ensures the high quality of the generated WSI virtual stain (refer to Figure.\ref{fig:demo_crohn}~and~\ref{fig:demo_kidney}) while offering the spatial context, keeping the processing time manageable, thus aligning with the needs and dynamics of modern anatomopathological practice.

\subsection{Trust in virtual stains through self-inspection--anomaly detection}

A significant barrier to integrating generative models, particularly in healthcare, is the absence of a confidence score with the generated output. This limitation raises critical questions: How can we detect problems in the input H\&E data? What is the model's confidence in its virtual stains? How can we assess the quality of the synthetic stains and identify errors that might influence a pathologist’s decision to rely on virtual stains or request traditional chemical stains for confirmation?

Such concerns are paramount in high-stakes decision-making processes. There is a pressing need for interpretable methods to ensure that these powerful generative approaches are not sidelined due to a lack of trustworthiness. In this study, we leverage knowledge-guided training not only to enhance control over the learning process—which has demonstrated improvements in performance (refer to Section~\ref{subsec:L_he+L_ihc})—but also to provide pathologists with an interpretable narrative. This approach considers stain masks, shifting the model's focus to medically relevant features, thus providing a clearer explanation than a fully black-box model.

Moreover, we utilize the discriminator’s learned knowledge from the training phase—an element often discarded post-training—which models the authenticity of images. This unique application allows for the inspection of data quality and its deviation from the learned distribution.

\begin{figure}[H]
    \centering
    \includegraphics[width=0.9\linewidth]{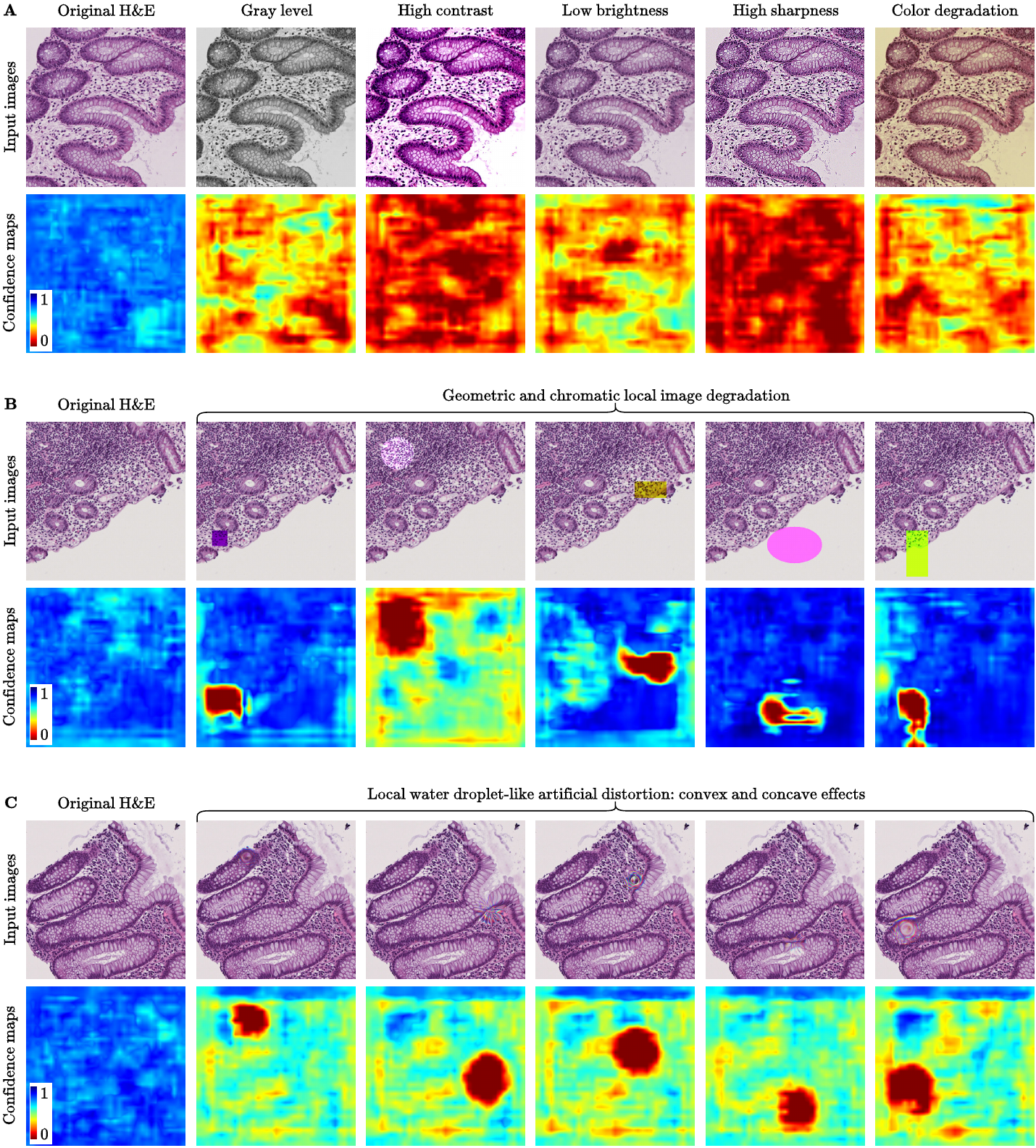}

    \caption{\textbf{Comparison of original and degraded H\&E images with corresponding H\&E discriminator's confidence maps}: Panels \textbf{A}, \textbf{B}, and \textbf{C} demonstrate the analysis of H\&E-stained tiles. In each panel, the top row displays the original H\&E tile alongside five degraded versions of the same tile, while the bottom row presents the associated discriminator's confidence maps. These maps highlight areas of perceptual inconsistency, marked in red. Panel \textbf{A} illustrates global degradation potentially caused by issues in chemical staining or scanning errors, such as incorrect staining concentration or scanner configuration problems, with the model successfully identifying such global defects. Panel \textbf{B} shows local contamination possibly due to chemical staining errors or physical artifacts on the scanner, with the model pinpointing the locations of the contamination. Panel \textbf{C} depicts artifacts resembling water droplets that can adhere to the slides during preparation, potentially causing analysis errors; here, the model indicates the positions of these droplet-like artifacts, thereby drawing expert attention to the affected regions. For reproducibility details, refer to Section~\ref{method:QC}.}
    \label{fig:patchgan_exp}
\end{figure}

To demonstrate the utility of this approach, we processed H\&E tiles and examined global degradation that could arise from incorrect stain concentration or scanner configurations, as illustrated in Figure~\ref{fig:patchgan_exp}. The discriminator effectively detects the domain shift in these degraded images, aligning with an anomaly detection framework. It flags deviations from the learned distribution in red, as shown in Figure~\ref{fig:patchgan_exp}.  These results support the hypothesis that H\&E tiles presenting new defects can be detected via the model's discriminator.

\begin{figure}[H]
    \centering
    \includegraphics[width=1\linewidth]{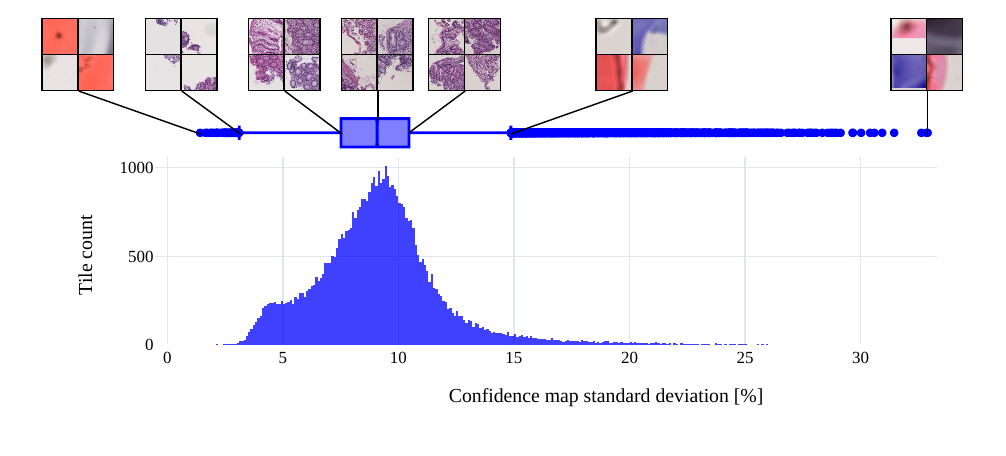}
    \caption{\textbf{Discriminator confidence mapping for H\&E tile authenticity evaluation (anomaly detection).} This figure evaluates the authenticity of 47984 H\&E-stained tiles from 2022 authentic WSIs (H\&E stained during 20 years time interval with different scanners) using discriminator confidence maps. The maps' standard deviation is used to assess each tile's authenticity. The histogram provides pathologists with an empirical tool to determine the acceptable H\&E range (e.g., 3.11\% to 14.86\%), identifying tiles within this range as highly authentic. Tiles outside this range are flagged as outliers, typically due to being background or significantly degraded, indicated by unusually high or consistently low deviations on the confidence maps. These results highlight the discriminator's capability to identify and quantify tile authenticity, serving as an essential tool for pathologists to exclude unreliable artifacts during the H\&E staining and scanning processes. This approach effectively prevents the introduction of substandard images into the multi-virtual staining pipeline, thereby reducing the potential error rate in synthetic stains and enhancing the reliability and trustworthiness of generated outputs. For reproducibility details, refer to Section~\ref{method:QC}.}
    \label{fig:QC_STD}
\end{figure}

An extensive evaluation was conducted on 2022 authentic WSIs from a private dataset, comprising 47984 tiles (512x512 H\&E tiles). Given that anomalies can be local or global, we employed the standard deviation of the discriminator's confidence map as an indicator, as depicted in Figure.~\ref{fig:QC_STD}. This analysis not only confirms that the discriminator can detect outliers (e.g., artifact tiles, mostly background tiles) but also facilitates the empirical determination of a confidence interval (e.g.,~3.11\%<acceptable<14.86\%). This method introduce an effective filter to avoid feeding the multi-virtual staining approach with unsound H\&E images (garbage in garbage out), thus, reduces the potential error rate in synthetic stains, enhancing reliability and trust in generated outputs.

\begin{figure}
    \centering
    \includegraphics[width=\linewidth]{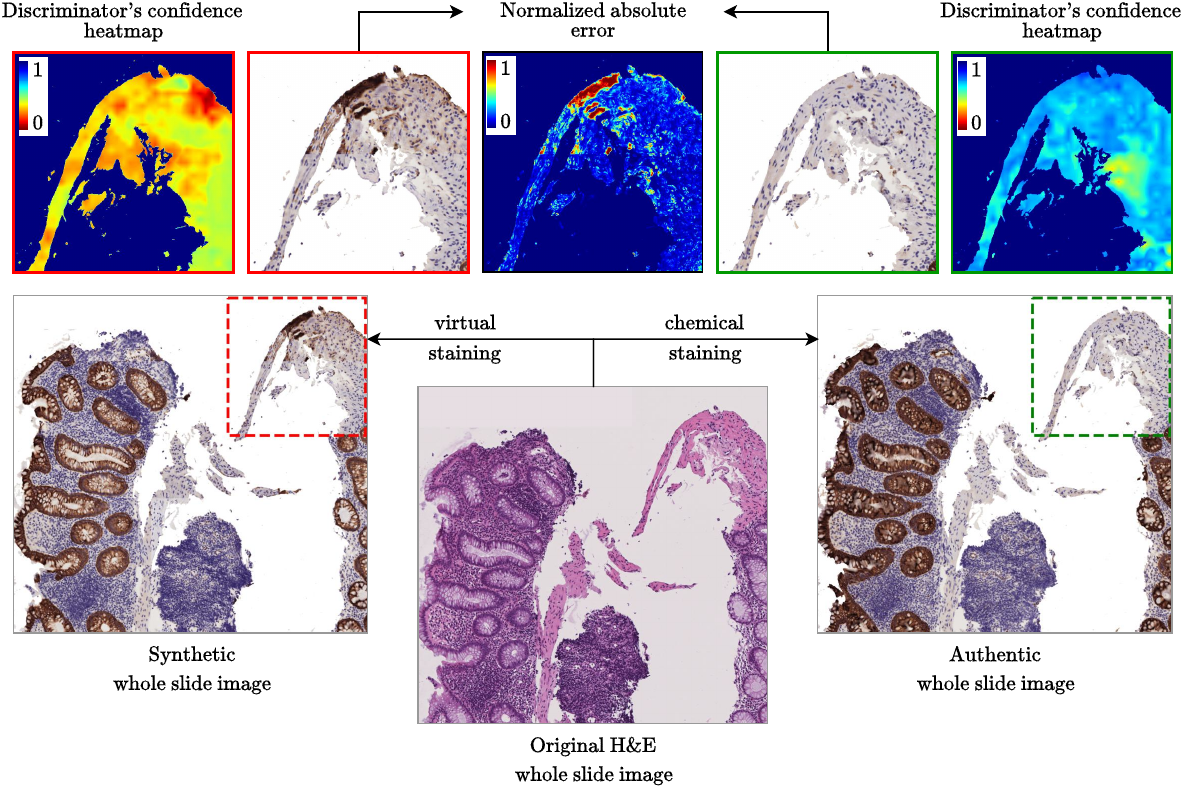}
    \caption{\textbf{Visualization of discriminator confidence in virtual staining analysis.} This figure illustrates the effectiveness of discriminator confidence maps in evaluating virtual and authentic stained WSIs. Two tissue sections are shown: one stained authentically and the other virtually stained with an identifiable error. The discriminator’s response is visualized through heat maps, where areas of discrepancy are highlighted in red. These highlighted regions correspond to significant deviations from the expected stain appearance, providing pathologists with a pixel-wise confidence score. This visualization aids in determining the necessity of additional confirmatory chemical staining and in identifying critical areas for detailed examination. By quantifying and displaying errors, this tool reinforces the reliability of virtual staining technologies and supports pathologists in making more informed decisions. For reproducibility details, refer to Section~\ref{method:QC}.}
    \label{fig:disc-XAI}
\end{figure}

In addition, we employed the discriminator's confidence maps on the outputted virtual stains to generate pixel-wise confidence scores. These scores empower pathologists by highlighting regions where the virtual staining deviates from the expected representation of a stain. This feature acts as a secondary filter in the output stage of our pipeline, visually represented through heat maps as depicted in Figure~\ref{fig:disc-XAI}. The figure contrasts the discriminator's responses to identical tissue sections—one from an authentically stained WSI and the other from a virtually stained WSI with a staining error. Notably, the discriminator pinpoints staining discrepancies in red, aligning with the actual differences computed between the authentic and virtual images. This methodology demonstrates the approach's capacity to provide pathologists with a reliable confidence score, offering additional context to determine the significance of the region in question for specific use cases. It also assesses whether it is necessary to perform a chemical stain to confirm findings, thus eliminating any uncertainties. By clearly indicating areas of uncertainty in the output, this tool builds greater trust in virtual staining technologies, reassuring users of its reliability and enhancing overall confidence in the outputs provided.

This study highlights the quality-check capability of integrating discriminator confidence maps into the workflow of digital and virtual staining in pathology (refer to Figure~\ref{fig:virtual-XAIinfrence}). Our approach's ability to identify discrepancies and artifacts at both input and output stages ensures that only high-quality, reliable data are utilized and generated, addressing the critical issue of "garbage in, garbage out" in medical imaging. Looking forward, the adoption of such advanced tools promises to refine the precision of digital pathology, potentially leading to more personalized and timely therapeutic interventions.

\subsection{Cloud-based digital pathology for enhanced efficiency and usability--proof-of-concept}

Configuring the software and hardware required for complex generative models can be both time-consuming and resource-intensive, demanding specific technical skills that may not be readily available in the busy environments typical of pathology labs. An anywhere-accessible system, preferably through a browser, could significantly enhance time efficiency and work comfort. In this study, our goal is to provide a holistic approach to managing multi-virtual staining. Therefore, we have chosen to use Cytomine~\cite{cytomine}, an open-source platform, as a proof of concept to deploy our multi-virtual staining technique (discussed in Section~\ref{sec:cloud_sota}). This choice allows us to bridge the gap between cutting-edge DL models and their day-to-day application, offering replicable guidelines for utilizing open-source, cloud-based platforms.

\begin{figure}
    \centering
    \includegraphics[width=1\linewidth]{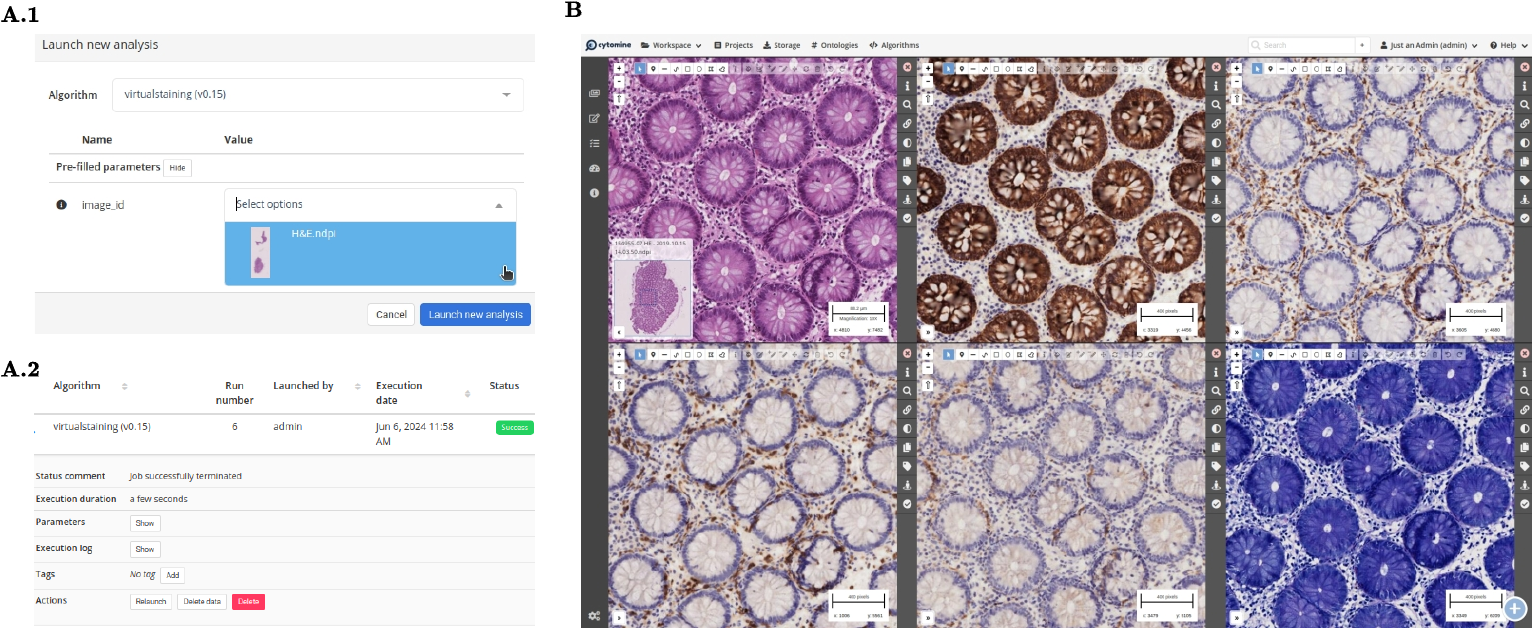}
    \caption{\textbf{Towards effortless digital digital histopathology through cloud-based multi-virtual staining: Proof-of-concept.} \textbf{A.1.} displays a user interface for selecting the desired H\&E WSI and setting the parameters for inference. \textbf{A.2.} illustrates the panel that tracks the progress of the multi-virtual staining process (slurm job). \textbf{B.} presents synchronized views of an array of virtually stained slides alongside the original H\&E slide (upper left). This figure demonstrates our dockerized multi-virtual staining implementation on the open-source Cytomine platform\cite{cytomine} as a use-case. Computations are performed on a backend server (via slurm), with the user only required to upload the H\&E slide and initiate the algorithm through the browser. Results are then displayed in a synchronized view, significantly minimizing user effort. For reproducibility details, refer to Section~\ref{method:cloud}.}
    \label{fig:cytomine}
\end{figure}

To achieve this, we deployed the platform and dockerized our multi-virtual staining implementation before integrating it into the cloud-based platform. This approach enables pathologists to easily execute complex algorithms directly through their web browser, as illustrated in Figure~\ref{fig:cytomine}. This streamlined integration simplifies the use of advanced DL models in routine pathological analysis, enhancing the accessibility and practicality of digital histopathology tools.

\subsection{Paired H\&E-multi-stains dataset in the context of pediatric Crohn’s disease at diagnosis}

As detailed in Section~\ref{sec:data_sota}, one primary challenge in multi-stain data analysis is the scarcity of publicly available datasets. For instance, the dataset from de Haan et al. (2021) was not shared~\cite{deHaan2021}. Additionally, the availability of high-quality paired data is limited; typically, datasets such as AHNIR~\cite{AHNIR} are compiled from adjacent slides, resulting in imperfectly matched samples. Specifically, the AHNIR kidney dataset contains only a limited set of slides, with five slides for each stain type: H\&E, PAS, PASM, and MAS.

This issue is prevalent in other studies as well; for example, MVFStain~\cite{ZHANG2022MVFStain} utilized only a fraction of the AHNIR dataset for lung lesions, employing one WSI for training and another for testing. Similar methodologies are applied to datasets concerning lung lobes and breast tissues, utilizing two WSIs for training and one for testing to maintain methodological consistency.

These challenges hinder not only the public availability of such data but also limit the diversity and quality of the datasets. With samples derived from adjacent slides, significant pairing challenges arise, complicating the objective evaluation of computational methods. This necessitates the use of elastic registration (e.g. VALIS~\cite{valis}), which is error-prone due to variations in tissue characteristics.

\begin{figure}[H]
    \centering
    \begin{subfigure}{0.24\textwidth}
        \includegraphics[width=\linewidth]{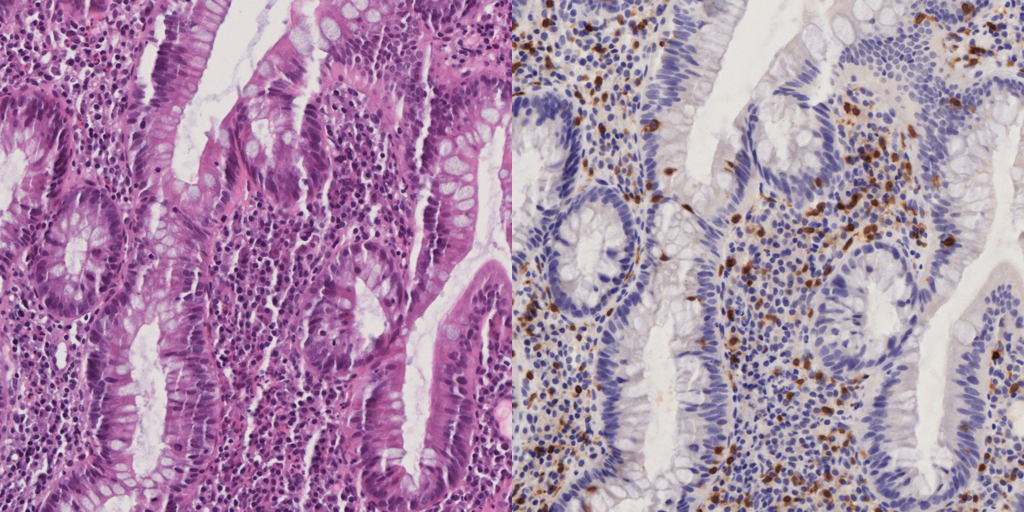}
        \caption{H\&E $\leftrightarrow$ CD3}
    \end{subfigure}
    \hfill
    \begin{subfigure}{0.24\textwidth}
        \includegraphics[width=\linewidth]{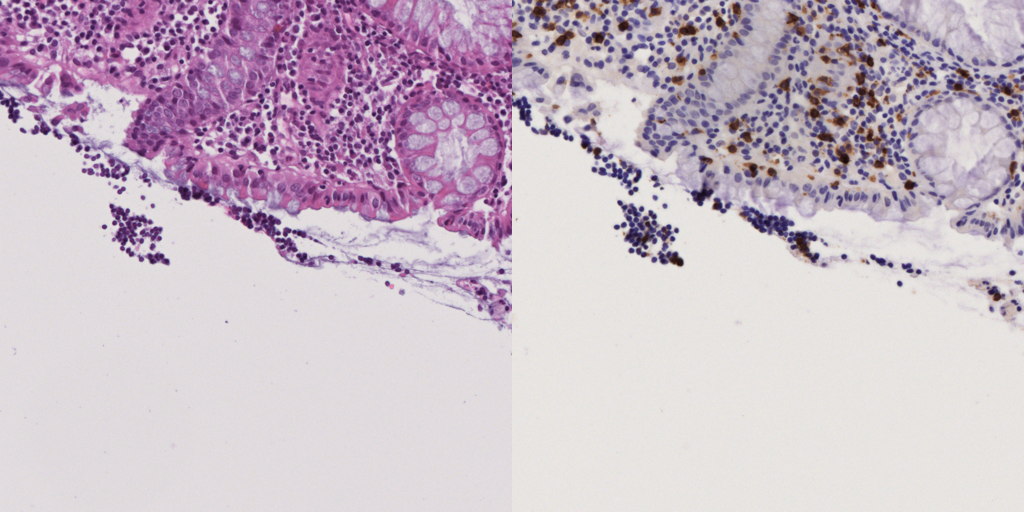}
        \caption{H\&E $\leftrightarrow$ CD8}
    \end{subfigure}
    \hfill
    \begin{subfigure}{0.24\textwidth}
        \includegraphics[width=\linewidth]{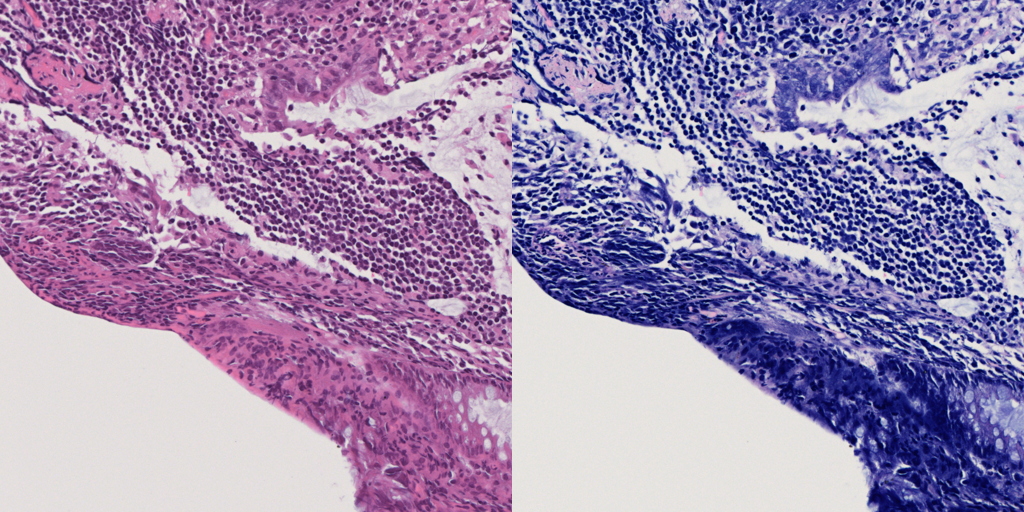}
        \caption{H\&E $\leftrightarrow$ GIEMSA}
    \end{subfigure}
    \hfill
    \begin{subfigure}{0.24\textwidth}
        \includegraphics[width=\linewidth]{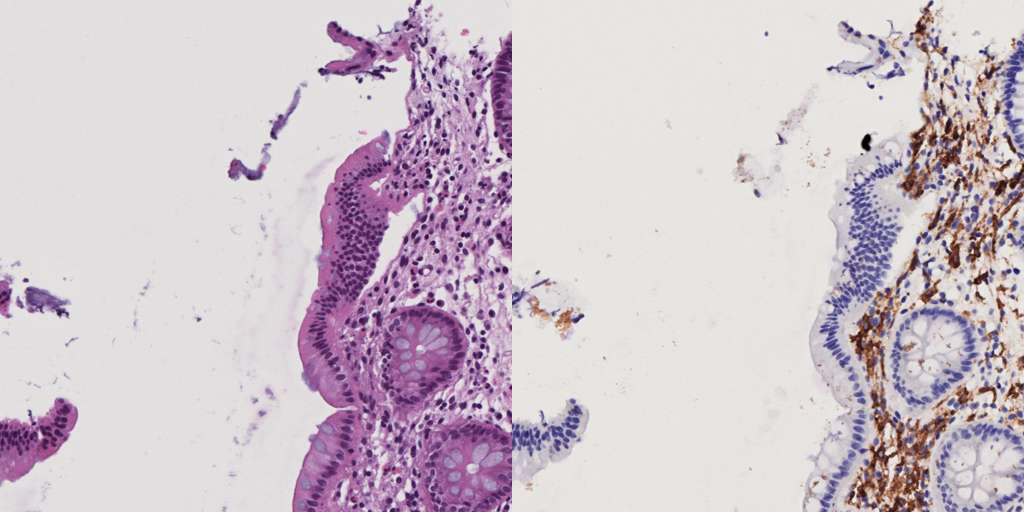}
        \caption{H\&E $\leftrightarrow$ CD163}
    \end{subfigure}

    \vspace{1em} 
    \begin{subfigure}{0.24\textwidth}
        \includegraphics[width=\linewidth]{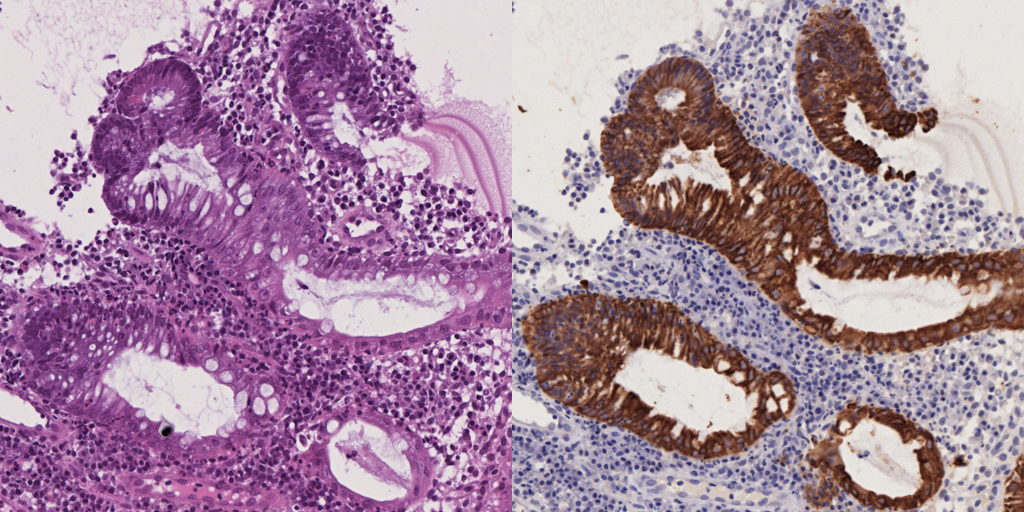}
        \caption{H\&E $\leftrightarrow$ AE1AE3}
    \end{subfigure}
    \hfill
    \begin{subfigure}{0.24\textwidth}
        \includegraphics[width=\linewidth]{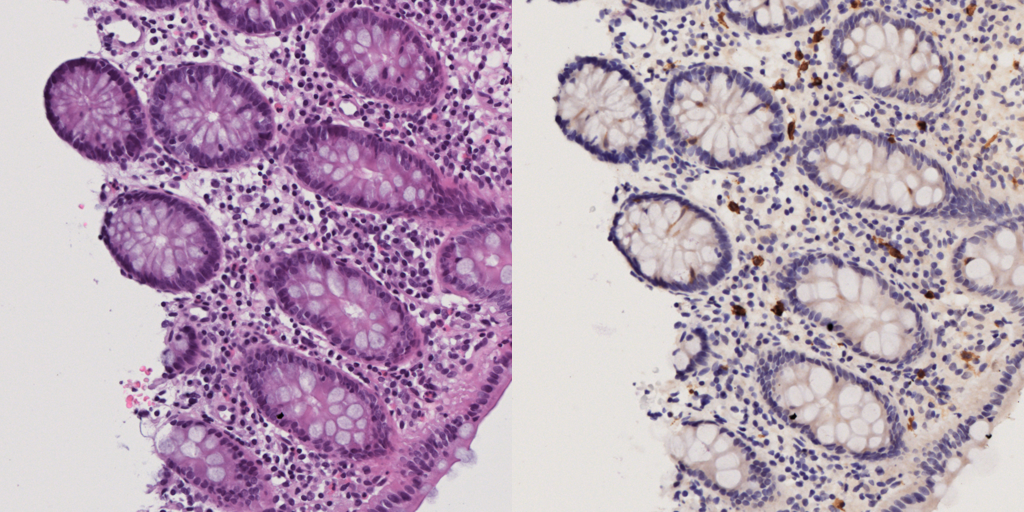}
        \caption{H\&E $\leftrightarrow$ CD117}
    \end{subfigure}
    \hfill
    \begin{subfigure}{0.24\textwidth}
        \includegraphics[width=\linewidth]{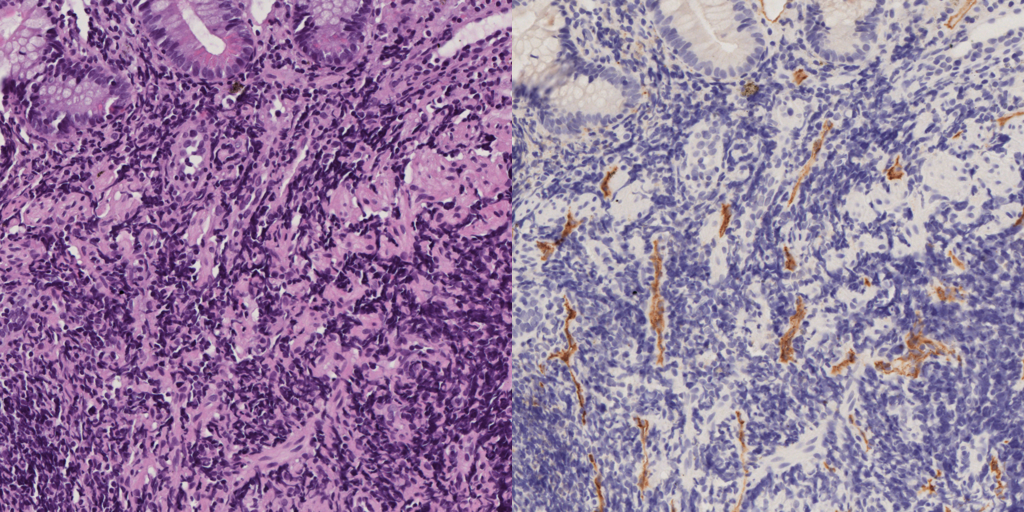}
        \caption{H\&E $\leftrightarrow$ D2-40}
    \end{subfigure}
    \hfill
    \begin{subfigure}{0.24\textwidth}
        \includegraphics[width=\linewidth]{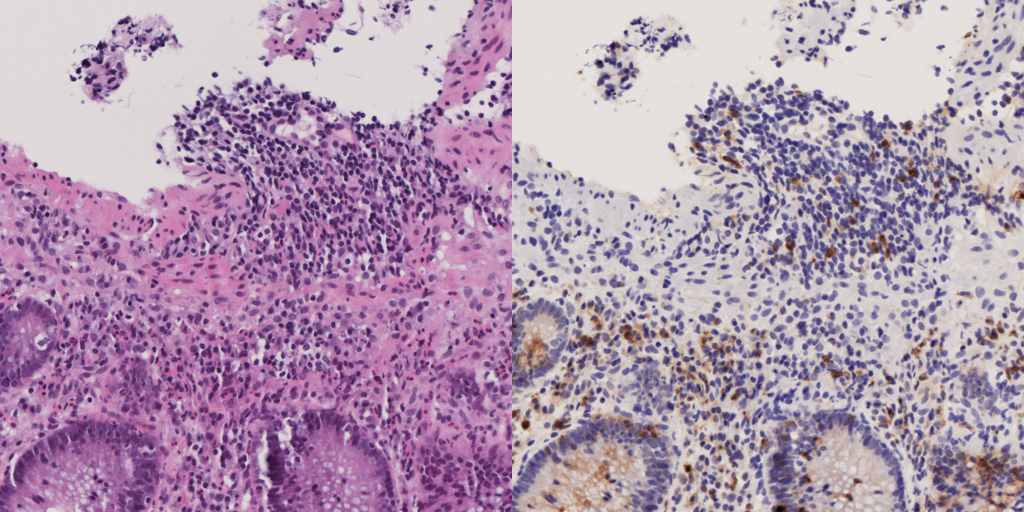}
        \caption{H\&E $\leftrightarrow$ CD15}
    \end{subfigure}

    \caption{\textbf{Samples visualization of the multi-stain pediatric Crohn's disease dataset, showcasing paired H\&E for different stain types.} This figure illustrates the perfect pairing of WSIs from identical tissue sections, which is central to the dataset's utility in computational pathology research.}\label{fig:crohn_data}
\end{figure}

To overcome these limitations, we propose the introduction of a new dataset that provides paired H\&E to eight different stains, focusing on pediatric Crohn's disease. This dataset aims to catalyze further research in computational pathology by including 30 H\&E WSIs and 30 stained WSIs across eight stains (AE1AE3, CD117, CD15, CD163, CD3, CD8, D240, and GIEMSA), culminating in a total of 480 WSIs. Each sample consists of perfectly matched data from identical tissue sections, as depicted in Figure~\ref{fig:crohn_data}. This comprehensive collection of 480 WSIs aims to drive advancements in computational pathology. By providing high-quality, diverse data, we anticipate setting a new benchmark for methodologies in not only virtual staining but also in segmentation, detection, and other computational histopathology applications.

\section*{Discussion}

Our investigation into the current state of the art in computational pathology has revealed critical challenges, notably the opaque nature of deep learning technologies and a shortage of high-quality public data. These issues significantly hinder the integration of advanced computational tools into routine clinical practice. To address these challenges, our study proposes a holistic approach centered on enhancing performance, trustworthiness, scalability, and data quality and quantity. This approach ensures that complex systems are accessible through secure, cloud-based platforms, which is crucial for their successful integration into the field.

The methodology we developed significantly contributes to computational pathology by improving scalability through compressive regularization and knowledge-guided methodologies during both training and inference phases. Trust is further enhanced by incorporating discriminators for input quality control and output confidence scoring. The practical implementation of our model in an open-source, cloud-based deployment for virtual staining demonstrates promising potential for real-world applications.

In advancing the field's understanding of virtual staining, we released a dataset of 480 whole slide images. This not only sets a new standard for quantitative evaluation in computational pathology but also supports diverse applications such as segmentation and detection. By making these resources available, we encourage the scientific community to engage in more reproducible research using this dataset.

Looking ahead, expanding our dataset to include a wider range of pathological conditions beyond pediatric Crohn's disease will enhance the generalization of our model. Further exploration of various deep learning architectures is expected to improve overall system performance.

In conclusion, our research introduces significant enhancements to computational pathology by integrating a unified H\&E encoder, adapted loss functions, regularization techniques, and context-driven learning within a cloud-based framework. These advancements not only meet but exceed current standards of quality and trustworthiness in stain transformations, paving the way for a more reliable, accessible, and effective future in computational pathology, ultimately contributing to better clinical outcomes.
\section{Methods}

Our methodology integrates deep learning techniques with innovative computational pathology strategies to enhance the scalability, accuracy, trustworthiness, and applicability of virtual stain transformations. In the following sections, we provide comprehensive details about our data, the deep learning model used, and the training approaches employed.

Addressing the limitations of virtual staining, as discussed in Section~\ref{sec:stain_SOA}, our methodology is guided by several foundational principles. At its core, we developed a single H\&E encoder that supports multiple stain generators. This configuration places increased emphasis on the H\&E encoder to precisely highlight and emphasize essential morphological regions within the tissue. Consequently, this design ensures a richer latent representation while reducing over-fitting due to increased task density, directly enhancing performance.

Moreover, during the training phase, our method enhances both performance and trust by computationally leveraging IHC-activated regions. By integrating knowledge from stains in a self-supervised manner and emphasizing stain generation-adapted loss functions and regularization, we produce medically accurate stains, thereby strengthening the credibility of the training process.

A distinctive feature of our approach is the deployment of discriminators during production to ensure quality in two main aspects. Firstly, the discriminators assess the quality of input H\&E WSIs pixel-wise, alerting users to potential data impurities and providing heatmaps as visual feedback through XAI methods. Secondly, they generate pixel-wise confidence scores for synthetic stains, which can also be visualized using XAI methods to produce heatmaps.

Enhancing trustworthiness further, we propose a compartmentalized design that offers unparalleled flexibility and practicality. Because different model components are separable, it facilitates the seamless incorporation of data quality check methods and visual XAI tools during production. Moreover, during deployment, it is feasible to load only the required parts of the model based on a pathologist's specific stain requirements, eliminating the need to load the entire model. Importantly, if a new stain type is introduced to the dataset, our design enables efficient training for this addition alone, bypassing the need for exhaustive model retraining. These features highlight the scalability and practicality of our approach.

To ensure user-friendliness and a seamless experience for pathologists, we offer a proof of concept of our system accessible through a web-based interface, with all computational tasks executed in the background (back-end) on a cloud-based platform.

\subsection{Data}\label{method:data}

In this study, we introduce a rigorously curated dataset that is crucial for our research. This dataset was acquired at Robert Debré Hospital in Paris within a study on pediatric Crohn's disease. This study has been approved by the INSERM ethic committee (IRB3888, ref 21-761) in March 2021.

\textbf{Population: } the study focuses on pediatric and adult patients diagnosed with Crohn's disease according to the ESPGHAN criteria (European Society for Paediatric Gastroenterology Hepatology and Nutrition). These patients were followed at Robert Debré Hospital for at least one year and had an initial biopsy at the time of diagnosis. The study includes all patients diagnosed at the center from 1988 to 2019. Patients for whom whole slides were too old were removed, resulting in one or multiple slides available for 59 patients. This population includes more male individuals (69\%) and has a mean age of 11.11 years (standard deviation 3.64).

\textbf{Dataset description: } the dataset comprises 480 digital slides, evenly distributed across eight paired combinations of H\&E and IHC stains. Each combination includes 30 matched pairs of an H\&E slide and its IHC-stained counterpart, featuring the following markers: Anticytokeratin AE1/AE3 (AE1AE3), CD117 (c-Kit), CD15 (Lewis X or SSEA-1), CD163 (macrophages marker), CD3 (T-cell co-receptor), CD8 (T-cell co-receptor), Cluster D2-40 (D240), and Giemsa stain. To maintain consistency and reliability, all slides were scanned at a uniform magnification of 40x using the same scanner with a resolution of 0.22$\mu m$ per pixel. For the experiments, the dataset's slides are randomly divided into two subsets: 20\% for testing and the remaining 80\% for training. This division accounts for the tissue quantity per slide, specifically considering the number of tiles (at least 10\% of tissue) in each slide (refer to Section~\ref{sup:crohn_data} for exact slide IDs used for testing).

\subsection{Multi-virtual staining model architecture and training methodology}
\label{sec:arch_DL}

Our research primarily concentrates on adapting two notable neural network architectures, ComboGAN\cite{ComboGAN2017} and CycleGAN\cite{cycleGAN}, for a specific application: transforming H\&E-stained histological slides into various other stain types. The core architectural framework of our approach includes key components such as an encoder \(E_i\), a generator \(G_i\), and a discriminator \(D_i\), where \( i \) denotes the index representing each unique stain type within the set \{1,..., S\}. A pivotal contribution of our method is the utilization of shared (unique) H\&E-specific encoder, generator, and discriminator across all \( S \) stains, denoted as \( E_{\text{H\&E}} \), \( G_{\text{H\&E}} \), and \( D_{\text{H\&E}} \) respectively. This strategic choice is aimed at enhancing the efficacy and specificity of the transformation process.

\textbf{Synthesis training:} The training methodology we employ is based on a dual-cycle process aimed at transforming and reconstructing stains in histopathology slides. This approach focuses on seamlessly converting between H\&E stained tiles and various other stain types, while ensuring preservation of core structural features across transformations.

In the first cycle of this process "H\&E cycle", we start with an H\&E-stained tile (refer to Fig.~\ref{fig:train_unpaired_IHC}.A). This tile is first processed by an encoder specific to H\&E stains, denoted as \(E_{\text{H\&E}}\), transforming it into a latent representation \(Z_{\text{H\&E}}\). This latent representation is then fed into a generator \(G_i\), corresponding to the target stain type \(i\), producing an image \(\hat{Y}_i\) that mirrors the characteristics of the desired stain. To complete the cycle, this generated image is passed through another encoder \(E_i\), obtaining a new latent representation \(\hat{Z}_{\text{H\&E}}\), which is subsequently used by the H\&E generator \(G_{\text{H\&E}}\) to recreate an H\&E-stained tile \(\hat{X}_{\text{H\&E}}\). Mathematically, this re-conversion process, ensuring the cycle from an H\&E stain to a target stain \(i\) and back to H\&E, can be encapsulated as follows:

\begin{align}
\hat{Y}_i &= G_i\left(E_{\text{H\&E}}\left(X_{\text{H\&E}}\right)\right), \\
\hat{X}_{\text{H\&E}} &= G_{\text{H\&E}}\left(E_i\left(\hat{Y}_i\right)\right) \quad \forall i \in \{1, \ldots, S\}
\end{align}

In the second cycle "stain \(i\) cycle", we address the reverse process: starting with a tile originally stained with a specific type \(i\), we aim to convert it into an H\&E-stained representation and subsequently revert it to its original stain (refer to Fig.\ref{fig:train_unpaired_IHC}.B). Initially, the tile \(X_i\) undergoes encoding via \(E_i\) to produce a latent representation \(Z_i\). This is then converted into \(\hat{Y}_{\text{H\&E}}\) an H\&E-stained tile by the H\&E generator \(G_{\text{H\&E}}\), effectively translating \(X_i\) into the H\&E domain. The resulting H\&E image is re-encoded by \(E_{\text{H\&E}}\) into a new latent representation \(\hat{Z}_i\), which serves as input for the generator \(G_i\), reconstructing the original stained image \(\hat{X}_i\). This enables the conversion of any specific stain type to an H\&E representation and back, formulated mathematically as:

\begin{align}
\hat{Y}_{\text{H\&E}} &= G_{\text{H\&E}}\left(E_i\left(X_i\right)\right), \\
\hat{X}_i &= G_i\left(E_{\text{H\&E}}\left(\hat{Y}_{\text{H\&E}}\right)\right) \quad \forall i \in \{1, \ldots, S\}
\end{align}

To effectively train our architecture, we introduce a comprehensive global synthesis loss, which we denote as \(\mathcal{L}_{\text{IHC}, i}\). This loss facilitates the translation from H\&E stained images to a specific target stain, represented by \(i\). The process involves comparing the reconstructed image \(\hat{X}_i\) and the target image \(X_i\) for the specified stain \(i\), as well as comparing the reconstructed H\&E image \(\hat{X}_{\text{H\&E}}\) and the original H\&E image \(X_{\text{H\&E}}\). These comparisons are used in estimating the cycle loss for each stain \(i\), denoted as \(\mathcal{L}_{\text{cyc},i}\), which quantifies the fidelity of the translation between the H\&E stain and stain \(i\) in a bidirectional manner.

In addition to the cycle loss, our model also incorporates an adversarial loss, \(\mathcal{L}_{\text{adv}, i}\). This component utilizes discriminators in inference, specifically \(D_{\text{H\&E}}\) for the H\&E stain and \(D_i\) for the target stain \(i\), to assess the authenticity of the generated images \(\hat{Y}_i\) and \(\hat{Y}_{\text{H\&E}}\). The discriminators aim to distinguish between real and synthesized images, thus encouraging the generation of images that are indistinguishable from genuine stained samples. Additionally, the model incorporates a regularization loss, \(\mathcal{L}_{\text{reg}, i}\), which aids in the convergence of the model. The overall synthesis loss function for each specific stain \(i\), out of a total of \(S\) stains, is given by the equation:

\begin{equation}
\mathcal{L}_{\text{IHC}, i} = \lambda_{\text{cyc}}  \cdot \mathcal{L}_{\text{cyc}, i} + \lambda_{\text{adv}} \cdot \mathcal{L}_{\text{adv}, i} +\lambda_{\text{reg}} \cdot \mathcal{L}_{\text{reg}, i} \quad \forall i \in \{1,..., S\}
\end{equation}

Here, \(\lambda_{\text{cyc}}\), \(\lambda_{\text{adv}}\) and \(\lambda_{\text{reg}}\) are weighting coefficients. This approach ensures a versatile and effective training regime that can accommodate a range of stains, denoted by \(S\), enhancing the model's ability to generalize across a wide range of stains.

\textbf{Discriminator training:} In our approach, we employ two types of discriminators: one for the H\&E stains, denoted as \( D_{\text{H\&E}} \), and individual discriminators for each IHC stain, represented as \( D_i \) for the \(i^{th}\) stain. Each discriminator is trained using its respective loss function to optimize its performance. 

For the H\&E discriminator, \( D_{\text{H\&E}} \), the loss function is defined as follows:

\begin{equation}
\mathcal{L}_{D_{\text{H\&E}}} = \lambda_{\text{D}} \cdot (\mathcal{L}_{\text{real}_{\text{H\&E}}} + \mathcal{L}_{\text{synthetic}_{\text{H\&E}}})
\end{equation}

This equation represents a combination of the losses \(\mathcal{L}_{\text{real}_{\text{H\&E}}}\) and \(\mathcal{L}_{\text{synthetic}_{\text{H\&E}}}\) from real and synthetic H\&E stained images respectively (refer to equation \eqref{eq:h&e_real_synthetic}), scaled by a weighting factor \( \lambda_{\text{D}} \), to measure the discriminator's performance in distinguishing between genuine and artificially generated H\&E images.

Similarly, for each stain discriminator \( D_i \), the loss function is tailored to assess its ability to discern real from synthetic stained images of that particular type, defined as:

\begin{equation}
\mathcal{L}_{D_{\text{IHC},i}} = \lambda_{\text{D}} \cdot (\mathcal{L}_{\text{real}, i} + \mathcal{L}_{\text{synthetic}, i}) \quad \forall i \in \{1, \ldots, S\}
\end{equation}

Here, \( \mathcal{L}_{\text{real}, i} \) and \( \mathcal{L}_{\text{synthetic}, i} \) correspond to the losses from real and synthetic images of the \(i^{th}\) stain, respectively (refer to equation \eqref{eq:ihc_real_synthetic}). The sum of these losses, weighted by \( \lambda_{\text{D}} \), constitutes the total loss for that discriminator, ensuring it effectively learns to differentiate between actual and generated samples of its specific stain. This structure is applied across all \( S \) stains, enabling the discriminators to specialize in their respective stains for improved performance in identifying authentic versus generated images.

\subsubsection{Integrating annotation free knowledge through loss function optimization}\label{method:Lcyc_Lihc}

In the development of computational models for synthesizing IHC slides, it has become evident that traditional distances such as \( \mathcal{L}_1\), \( \mathcal{L}_2\), and mean square error (MSE) pose significant challenges. A key issue with these metrics lies in their indiscriminate treatment of different regions on the slides, as they fail to differentiate between tissue sections and areas activated by IHC staining. This becomes particularly problematic due to the inherent staining imbalance in IHC slides, where the vast majority of the slide is IHC-negative, underscoring that IHC staining is specific to only a small fraction of the total slide content.

To address these shortcomings, our approach harnesses the cycle consistency loss \( \mathcal{L}_{\text{cyc}} \) and adversarial loss \( \mathcal{L}_{\text{adv}} \)~\cite{cycleGAN, ComboGAN2017}. Building on these foundations, we propose a novel method that integrates IHC-activated areas into the model's training process for synthesizing \(S\) stains from H\&E stained slides. This strategy facilitates the extraction of annotation-free knowledge, seamlessly incorporating the distinctive characteristics of each stain into the model. Consequently, our model can generate features with enhanced fidelity, thereby improving the overall quality of synthesis. By guiding the synthesis process with stain-specific insights, our method not only enhances performance but also enhances the transparency and trustworthiness of the synthesized images.

\begin{figure}[ht]
    \centering

    \begin{subfigure}{0.5\textwidth}
        \centering
        \includegraphics[width=\textwidth]{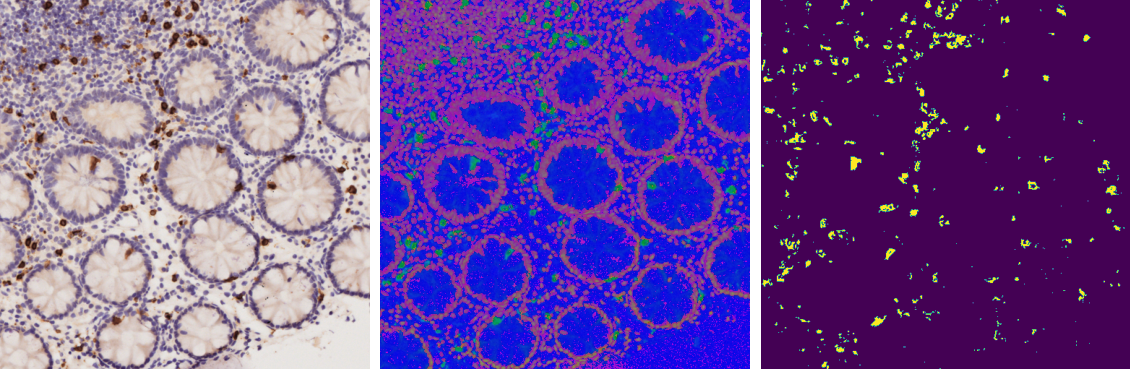}
        \includegraphics[width=\textwidth]{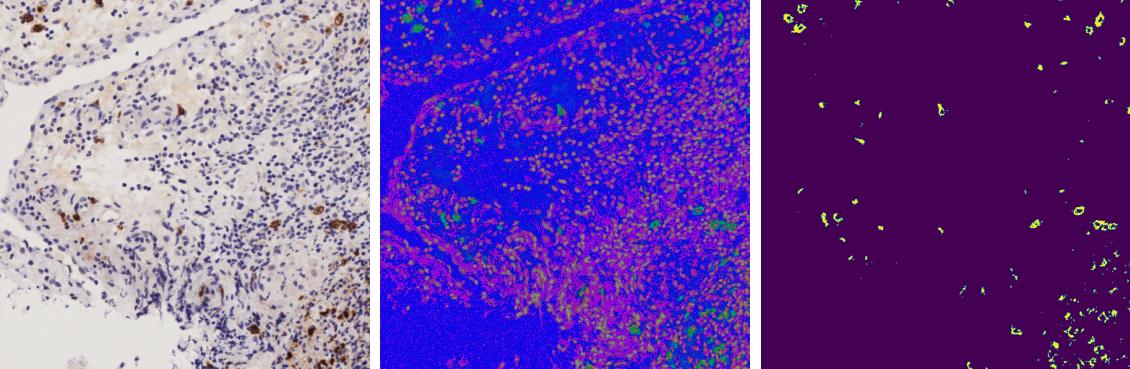}
        \includegraphics[width=\textwidth]{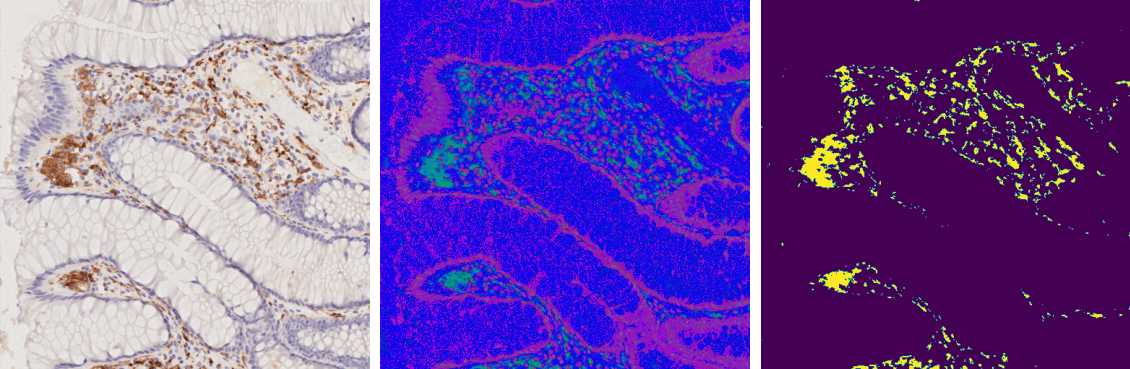}
    \end{subfigure}
    
    \caption{\textbf{Automated Extraction of IHC-Activated Regions from Stained Tiles:} For each instance, such as (from the top) CD8, CD117, CD163, the extraction process is visualized in a three-column format. The left column displays the original RGB stained tile (\( X_i \)); the middle column depicts the tile's conversion into the HSV color space, capturing the unique chromatic signature from antibody-tissue reactions; and the right column showcases the resulting binary mask (\( M_i \)) highlighted in yellow.}
    \label{fig:ihc_masks}
\end{figure}

To automate the extraction of a precise mask for the IHC-activated region, denoted as \( M_i \), from the \( i^{th} \) IHC-stained image, \( X_i \), we begin by transforming \( X_i \) from its original RGB color space to the HSV color space. This conversion is crucial as it significantly enhances the ability to isolate the target regions based on their color and brightness attributes. After converting the image to the HSV color space, we apply a threshold to isolate a distinct mask, \( M_i \), which is shown in Fig.\ref{fig:ihc_masks}. This mask \( M_i \) is used in calculating dynamic weighting factors, \( \alpha_{i} \) and \( \beta_{i} \), which are integrated into the computation of the loss function. These factors are defined as follows:

\begin{equation}
    \alpha_{i} = \frac{\Bar{N}_i}{N_i + \Bar{N}_i}, \quad \beta_{i} = \frac{N_i}{N_i + \Bar{N}_i}
\end{equation}

In this equation, \( N_i \) represents the total number of pixels within the foreground, which corresponds to the IHC-activated regions, and \( \Bar{N}_i \) represents the total number of pixels within the background, or the IHC-non-activated regions, of the \( i^{th} \) IHC-stained image. This approach enables a nuanced differentiation between the regions of interest and their background, facilitating more accurate analyses.

\textbf{Integrating knowledge during the synthesis training phase (cycle loss):}
Integrating knowledge during the training phase of synthesis involves a detailed process that utilizes both H\&E stained tiles and IHC images from specific stains. This process is important for creating high-quality, accurate synthetic images that mirror the unique characteristics of each stain.

In scenarios where direct pairs of H\&E-stained images and IHC images are not available, an \textit{unpaired setting} approach is used. This employs a specifically designed cycle loss, \(\mathcal{L}_{\text{cyc},i}\), to facilitate the synthesis in the absence of paired images. The cycle loss formula for the unpaired setting is:

\begin{equation}
\begin{split}
    \mathcal{L}_{\text{cyc},i} & = \mathcal{L}(\hat{X}_i, X_i) + \mathcal{L}(\hat{X}_{\text{H\&E}}, X_{\text{H\&E}}) \\
    & + \alpha_i\cdot\mathcal{L}(M_i \odot \hat{X}_i, M_i \odot X_i) + \beta_i\cdot \mathcal{L}(\bar{M}_i \odot \hat{X}_i, \bar{M}_i \odot X_i) \quad \forall i \in \{1, \ldots, S\}
\end{split}
\end{equation}

This approach meticulously focuses on the differentiation between IHC-activated regions (\(M_i\)) and non-activated regions (\(\bar{M}_i\)), ensuring the integrity of the synthesis process despite the absence of direct image correspondences (refer to Fig.\ref{fig:train_unpaired_IHC}).

Conversely, in \textit{paired settings} where direct correspondences between H\&E and IHC images exist, the cycle loss is formulated to ensure both the overall fidelity of image reconstructions and the accurate replication of specific IHC-activated regions while taking advantage of the direct correspondence (refer to Fig.\ref{fig:train_paired_IHC}). The comprehensive equation for the paired setting is:

\begin{equation}
\begin{split}
    \mathcal{L}_{\text{cyc},i} & = \mathcal{L}(\hat{X}_i, X_i) + \mathcal{L}(\hat{X}_{\text{H\&E}}, X_{\text{H\&E}}) \\
    & + \alpha_i\cdot[\mathcal{L}(M_i \odot \hat{X}_i, M_i \odot X_i) + \mathcal{L}(M_i \odot \hat{X}_{\text{H\&E}}, M_i \odot X_{\text{H\&E}})]\\
    & + \beta_i\cdot [\mathcal{L}(\bar{M}_i \odot \hat{X}_i, \bar{M}_i \odot X_i) + \mathcal{L}(\bar{M}_i \odot \hat{X}_{\text{H\&E}}, \bar{M}_i \odot X_{\text{H\&E}})]    \quad \forall i \in \{1, \ldots, S\}
    \label{eq:p_cyc}
\end{split}
\end{equation}

Both settings employ dynamic weighting factors, \( \alpha_i \) and \( \beta_i \), determined by the ratio of IHC-activated regions to non-activated regions within each image. This careful consideration ensures that the model's training prioritizes not only the overall accuracy of stain transformation but also the faithful replication of regions crucial for IHC analysis. Furthermore, the methodology incorporates masks \( M_i \) and \( \bar{M}_i \) in the computation of the cycle loss, enhancing the model's capacity to embed the distinctive characteristics of each stain directly into its architecture. This meticulous approach facilitates the integration of annotation-free knowledge, resulting in the production of high-quality, reliable synthetic images that accurately reproduce the complexities of IHC staining.

\textbf{Integrating knowledge during the synthesis training phase (adversarial loss):} In the context of the unpaired setting, the adversarial loss, denoted as \(\mathcal{L}_{\text{adv},i}\), is calculated by evaluating the authenticity of the generated images \(\hat{Y}_i\) and \(\hat{H\&E}\) through discriminators \(D_i\) and \(D_{\text{H\&E}}\), respectively. This process is enhanced by leveraging the stain mask \(M_i\), which assists \(D_{\text{H\&E}}\) in focusing on the areas activated by the IHC, as depicted in Fig.\ref{fig:train_unpaired_IHC}. The formula for computing adversarial loss in unpaired settings is as follows:
\begin{equation}
\begin{split}
    \mathcal{L}_{\text{adv},i} & = \mathcal{L}(D_i(\hat{Y}_i),1) \\
    & + \alpha_i\cdot\mathcal{L}(M_i \odot D_{\text{H\&E}}(\hat{Y}_{\text{H\&E}}), 1) \\
    & + \beta_i\cdot \mathcal{L}(\bar{M}_i \odot D_{\text{H\&E}}(\hat{Y}_{\text{H\&E}}),1) \quad \forall i \in \{1, \ldots, S\}
\end{split}
\end{equation}

For paired setting, the approach mirrors that of the unpaired setting but with an added emphasis on the direct correspondence between the \(H\&E\) stain and image \(i\), as shown in Fig.\ref{fig:train_paired_IHC}. The computation of adversarial loss for paired settings incorporates this direct correspondence and is expressed by the following equation:
\begin{equation}
\begin{split}
    \mathcal{L}_{\text{adv},i} & = \mathcal{L}(D_{\text{H\&E}}(\hat{Y}_{\text{H\&E}}),1) \\
    & + \alpha_i\cdot[\mathcal{L}(M_i \odot D_{\text{H\&E}}(\hat{Y}_{\text{H\&E}}), 1) + \mathcal{L}(M_i \odot D_i(\hat{Y}_i), 1)] \\
    & + \beta_i\cdot [\mathcal{L}(\bar{M}_i \odot D_{\text{H\&E}}(\hat{Y}_{\text{H\&E}}),1) + \mathcal{L}(\bar{M}_i \odot D_i(\hat{Y}_i),1) ]\quad \forall i \in \{1, \ldots, S\}
    \label{eq:p_adv}
\end{split}
\end{equation}

\textbf{Direct supervision loss (paired setting):} In the paired setting of our framework, the synthesis loss is meticulously designed to include a variety of components, notably the supervised loss (\(\mathcal{L}_{\text{sup},i}\)), alongside the cycle consistency loss (\(\mathcal{L}_{\text{cyc},i}\)) as referenced in equation \(\eqref{eq:p_cyc}\) and the adversarial loss (\(\mathcal{L}_{\text{adv},i}\)) as detailed in equation \(\eqref{eq:p_adv}\). Crucially, the supervised loss establishes a direct connection between the H\&E cycle and the specific stain cycle \(i\). It does this by evaluating the fidelity of the generated stains \(\hat{Y}_i\) and H\&E images \(\hat{Y}_{\text{H\&E}}\) against their actual counterparts (\(X_i\) and \(X_{\text{H\&E}}\), respectively). Furthermore, it incorporates a common mask (\(M_i\)), derived from \(X_i\), to concentrate the loss computation on pertinent areas of the image. This ensures that the generated images maintain both structural and stylistic integrity in relation to the original samples. The formula for the supervised loss is articulated as follows:

\begin{equation}
\begin{split}
    \mathcal{L}_{\text{sup},i} & = \mathcal{L}(\hat{Y}_i, X_i) + \mathcal{L}(\hat{Y}_{\text{H\&E}}, X_{\text{H\&E}}) \\
    & + \alpha_i\cdot[\mathcal{L}(M_i \odot \hat{Y}_i, M_i \odot X_i) + \mathcal{L}(M_i \odot \hat{Y}_{\text{H\&E}}, M_i \odot X_{\text{H\&E}})]\\
    & + \beta_i\cdot [\mathcal{L}(\bar{M}_i \odot \hat{Y}_i, \bar{M}_i \odot X_i) + \mathcal{L}(\bar{M}_i \odot \hat{Y}_{\text{H\&E}}, \bar{M}_i \odot X_{\text{H\&E}})]    \quad \forall i \in \{1, \ldots, S\}
    \label{eq:P_sup}
\end{split}
\end{equation}


\textbf{Integrating knowledge during the discriminator training phase:} To effectively train the \(D_i\) discriminator, we employ authentic \(X_i\) samples to generate corresponding stain masks \(M_i\). These masks enable the discriminator to discern nuances within IHC-activated regions through the \(\mathcal{L}_{\text{real}_i}\) loss function. Concurrently, the discriminator is trained to recognize synthetic images \(\hat{Y}_i\) as inauthentic using the \(\mathcal{L}_{\text{synthetic}_i}\) loss function. The formulation of both loss functions is as follows:
\begin{equation}
\begin{cases}
    \mathcal{L}_{\text{real}_i}  =  \mathcal{L}(D_i(X_i), 1) + \alpha \cdot \mathcal{L}(M_i \odot D_i(X_i), 1) + \beta \cdot \mathcal{L}(\Bar{M}_i \odot D_i(X_i), 1)\\
    \mathcal{L}_{\text{synthetic}_i} = \mathcal{L}(D_i(\hat{Y}_i), 0) \label{eq:ihc_real_synthetic}
\end{cases}
\end{equation}

In a similar vein, for the \(D_{\text{H\&E}}\) discriminator, we define \(\mathcal{L}_{\text{real}_{\text{H\&E}}}\) and \(\mathcal{L}_{\text{synthetic}_{\text{H\&E}}}\) to respectively discern the authenticity of H\&E stained images and identify synthetic counterparts. These loss functions are described as follows:
\begin{equation}
\begin{cases}
    \mathcal{L}_{\text{real}_{\text{H\&E}}}  = \mathcal{L}(D_{\text{H\&E}}(X_{\text{H\&E}}), 1)\\
    \mathcal{L}_{\text{synthetic}_{\text{H\&E}}} = \mathcal{L}(D_{\text{H\&E}}(\hat{Y}_{\text{H\&E}}), 0)
    \label{eq:h&e_real_synthetic}
\end{cases}
\end{equation}

\subsubsection{Enhancing training using regularization}

\textbf{One H\&E representation to rule all the staining modalities:}\label{method:Lhe} The primary objective of our approach is to develop a universal H\&E encoder and generator capable of handling all staining modalities used in histology. Traditional methods often face challenges due to varying requirements for different stains during the training phase. Some stains are inherently more complex to replicate, leading to uneven learning progress and potential neglect of less dominant stains. To address this issue, we have implemented a novel regularization strategy. This approach ensures an even distribution of learning focus across all stains by the components of our H\&E encoder, thereby minimizing bias towards any specific stain.

The regularization process involves a systematic selection where stains are randomly chosen from a complete set identified by numbers ranging from 1 to \(S\), ensuring comprehensive coverage. Updates are then applied to the encoder \(E_i\) and generator \(G_i\) for each selected stain \(i\), alongside updates for the shared H\&E components (\(E_{\text{H\&E}}\) and \(G_{\text{H\&E}}\)), ensuring equitable representation of every stain throughout the training cycles. After cycling through all \(S\) stains in a random sequence, we calculate a mean synthetic loss across them using the equation:

\begin{equation}
\mathcal{L}_{\text{H\&E}} = \frac{1}{S} \sum_{i=1}^{S} \mathcal{L}_{\text{IHC}, i}
\end{equation}

This calculated loss, \(\mathcal{L}_{\text{H\&E}}\), specifically refines the H\&E \(E_{\text{H\&E}}\) and \(G_{\text{H\&E}}\), marking the completion of a training iteration. This methodology ensures equal attention to each stain and significantly enhances the model's versatility across various modalities. As a result, we achieve a more stable and scalable training process, thereby enhancing the model's overall effectiveness in dealing with a wide range of staining modalites.

\textbf{Stain synthesis regularization:}\label{method:stain_regs} 
We propose a comprehensive methodology designed to encapsulate the entire spectrum of considerations for regularization in virtual staining. By integrating critical knowledge through IHC-activated regions across the stain mask \(M_i\). This process involves the calculation of a regularized loss, denoted as \( \mathcal{L}_{\text{reg},i} \), formulated as a weighted sum of three principal loss functions: identity (\(\mathcal{L}_{\text{idt,}i}\)), latent (\(\mathcal{L}_{\text{lat,}i}\)), and forward (\(\mathcal{L}_{\text{fwd,}i}\)) losses.

For every stain index \(i\) within the set \(\{1,..., S\}\), the regularized loss is precisely defined as:

\begin{equation}
\mathcal{L}_{\text{reg,}i} = \lambda_{\text{idt}} \cdot \mathcal{L}_{\text{idt,}i} + \lambda_{\text{lat}} \cdot \mathcal{L}_{\text{lat,}i} + \lambda_{\text{fwd}} \cdot \mathcal{L}_{\text{fwd,}i}
\end{equation}

Here, the coefficients \( \lambda_{\text{idt}}, \lambda_{\text{lat}}, \) and \( \lambda_{\text{fwd}} \) represent the respective weights of each loss component within the cumulative regularized loss framework.

The identity loss (\(\mathcal{L}_{\text{idt}}\)) quantifies the deviation between the original and generated images, utilizing the same domain encoder and generator within an auto-encoder setup. This ensures the encoder captures sufficient features for directly reproducing the input image. This concept is further extended to include \(M_i\) as follows:

\begin{equation}
\mathcal{L}_{\text{idt},i} = 
\begin{cases}
    \begin{aligned}
        &\mathcal{L}(G_i(E_i(X_i)), X_i) \\
        &+ \alpha \cdot \mathcal{L}(M_i \odot G_i(E_i(X_i)), M_i \odot X_i) \\
        &+ \beta \cdot \mathcal{L}(\Bar{M}_i \odot G_i(E_i(X_i)), \Bar{M}_i \odot X_i) & \quad \forall i \in \{1, \ldots, S\},
    \end{aligned}
    & \text{for stain } i \text{ image } X_i, \\
    \mathcal{L}(G_{\text{H\&E}}(E_{\text{H\&E}}(X_{\text{H\&E}})), X_{\text{H\&E}}) & \text{for H\&E image } X_{\text{H\&E}}.
\end{cases}
\end{equation}

The latent loss (\(\mathcal{L}_{\text{lat}}\)) aims at mitigating disparities within the latent space, specifically capturing the variance between the latent representation and its reconstructed version. This facilitates the alignment of embeddings from both H\&E and stain \(i\) encoders, incorporating IHC-activated regions as follow:

\begin{equation}
\mathcal{L}_{\text{lat},i} = 
\begin{cases}
    \begin{aligned}
        &\mathcal{L}(\hat{Z}_i, Z_i) \\
        &+ \alpha \cdot \mathcal{L}(M_i \odot \hat{Z}_i, M_i \odot Z_i) \\
        &+ \beta \cdot \mathcal{L}(\Bar{M}_i \odot \hat{Z}_i, \Bar{M}_i \odot Z_i) & \quad \forall i \in \{1, \ldots, S\},
    \end{aligned}
    & \text{for stain } i \text{ embeddings } Z_i \text{ and } \hat{Z}_i, \\
    \mathcal{L}(\hat{Z}_{\text{H\&E}}, Z_{\text{H\&E}}) & \text{for H\&E embeddings } Z_{\text{H\&E}} \text{ and } \hat{Z}_{\text{H\&E}}.
\end{cases}
\end{equation}

Finally, the forward loss assesses the divergence between the degraded versions (represented using lower case) of the original images (\(X_i, X_{\text{H\&E}}\)) and their corresponding outputs (\(\hat{Y}_{\text{H\&E}}, \hat{Y}_i\)), specified as:

\begin{equation}
\mathcal{L}_{\text{fwd},i} = 
\begin{cases}
    \begin{aligned}
        &\mathcal{L}(\hat{y}_{\text{H\&E}}, x_i) \\
        &+ \alpha \cdot \mathcal{L}(m_i \odot \hat{y}_{\text{H\&E}}, m_i \odot x_i) \\
        &+ \beta \cdot \mathcal{L}(\Bar{m}_i \odot \hat{y}_{\text{H\&E}}, \Bar{m}_i \odot x_i) & \quad \forall i \in \{1, \ldots, S\},
    \end{aligned}
    & \text{for stain } i, \\
    \mathcal{L}(\hat{y}_i, x_{\text{H\&E}}) & \text{for H\&E}.
\end{cases}
\end{equation}

By incorporating three distinct loss functions (\(\mathcal{L}_{\text{idt},i}\), \(\mathcal{L}_{\text{lat},i}\), and \(\mathcal{L}_{\text{fwd},i}\)) and leveraging knowledge from IHC-activated regions, this regularization approach opens up a wide array of opportunities for finely tuned handling of task-specific challenges. This method is particularly designed for unpaired settings, where it significantly contributes to the nuanced management of such tasks. In paired settings, these approaches do not offer substantial benefits due to the direct correspondences between the H\&E and the \(S\) stains, which are already extensively addressed by supervised loss functions. Nevertheless, the theoretical potential of integrating these regularization strategies \(\mathcal{L}_{\text{reg, i}}\) suggests broader applicability beyond their initially intended contexts.

\subsection{Trust in virtual stains through self-inspection–anomaly detection}\label{method:QC}

As described in Section~\ref{sec:arch_DL}, our architecture incorporates encoder, decoder, and discriminator components inspired by CycleGAN~\cite{cycleGAN} and ComboGAN~\cite{ComboGAN2017}. Central to our methodology is the use of a PatchGAN discriminator~\cite{patchgan,pix2pix,cycleGAN}, which is explicitly designed to differentiate between synthetic and authentic images. This discriminator features dual heads: one focusing on luminance (outputting a confidence map \(C_{lum}\)) and the other on RGB space (outputting a confidence map \(C_{rgb}\)). Each map rates the authenticity of the image on a clamped scale from -1 ('anomaly') to 1 ('authentic'), and both maps are resized to match the size of the input image.

To simplify the analysis for pathologists and reduce cognitive load, we combine these two maps into a single map by calculating the pixel-wise minimum of \(C_{lum}\) and \(C_{rgb}\), resulting in \(C_{all}\). This combined confidence map \(C_{all}\) is then normalized to a range of 0 to 1, where 0 indicates an anomaly and 1 indicates authenticity. This map can be used to compute various metrics, such as the standard deviation shown in Figure~\ref{fig:QC_STD}. Additionally, we apply a Jet-color map using OpenCV version 4.9.0~\cite{opencv} to transform \(C_{all}\) into an 8-bit unsigned integer RGB confidence map, as illustrated in Figure~\ref{fig:patchgan_exp}.

This approach provides comprehensive confidence maps that can identify a broad range of anomalies, as demonstrated in Figures~\ref{fig:QC_STD} and \ref{fig:patchgan_exp}. Given that the same discriminator architecture is utilized across all stains, this methodology is applicable to all of them.

\subsection{Tile-stitching for clean virtual staining WSI generation using a Hamming window-based approach}\label{method:tile_stitch}
To address the inevitable stitching artifacts encountered during the reconstruction of synthetic WSIs, we applied a tailored image processing approach. Central to our methodology was the use of a two-dimensional (2D) Hamming window~\cite{hamming1,hamming2}, designed to smooth the transitions between adjacent image patches and mitigate edge effects.

The Hamming window, traditionally used in signal processing~\cite{hamming1,hamming2} to taper the signal edges, was adapted to two dimensions to suit the image patches. Each patch, representing a portion of the larger image, was processed through this window to ensure a gradual transition at the borders. With an overlap > 0, this was achieved by computing the outer product of a one-dimensional Hamming window with itself, thus creating a symmetrical 2D window \( w(x, y) \) for a patch of size \( M \times M \) is defined as:

\begin{equation}
    w(x, y) = 0.54 - 0.46 \cos\left(\frac{2\pi x}{M-1}\right) \cdot \left(0.54 - 0.46 \cos\left(\frac{2\pi y}{M-1}\right)\right)
\end{equation}

where \( x, y \) range from 0 to \( M-1 \). This results in a 2D Hamming window which reduces the pixel values towards the edges of each patch. This window was then applied across the three color channels of the image. Each image patch (across all RGB cahnnels) was element-wise multiplied by this matrix, reducing the intensity at the peripheries and thereby softening the boundaries between stitched patches. This operation is described by the following equation:

\begin{equation}
P_{\text{weighted}}(x, y) = P(x, y) \cdot w(x, y)
\end{equation}

Where \( P(x, y) \) is the original pixel value at coordinates \( (x, y) \) within the patch for a given color channel, and \( w(x, y) \) is the value from the 2D Hamming window at these coordinates.

Post application of the Hamming window, the weighted patches were summed to form the complete WSI. In regions where patches overlapped, pixel values from multiple patches were combined. To ensure uniformity, the accumulated weights of the patches were recorded and used to normalize the pixel values in these overlapping areas. This normalization process was crucial for maintaining consistent intensity across the WSI, preventing visual discontinuities that could hinder the quality of the synthetically stained WSIs. This methodology can be applied to any tile-based virtual staining approach to reconstruct a clean WSI output.

The final processed image was saved in pyramidal TIFF format, suitable for high-quality WSI. The processing pipeline was implemented using Python, utilizing libraries such as NumPy~\cite{numpy} v1.26.3 for numerical operations and PyVIPS~\cite{pyvips} v2.2.2 for image handling, ensuring efficient memory usage and scalability.

\subsection{Cloud-Based Platform}\label{method:cloud}

We use the open-source software Cytomine Community Edition Legacy 3.1.0 to transform our virtual staining model in a web application. It operates based on a containerized architecture using Docker, which facilitates the creation and deployment of Cytomine applications through various modules (applications, web UI, databases, nginx proxy, jobs management, etc.). The core component for implementation of deep learning based applications is a \textit{software} Docker container. 

Our python-based application performing virtual staining is itself Dockerized and uploaded to software\_router where it is transformed in a Singularity image. Our python-based application includes the code for virtual staining, as well as the code to import the input WSI and upload the output virtual stains in the database. For these two last tasks, we use the Cytomine Python API for communication between the \textit{software} container and the image database container.

The inputs to this python-based application are specified in a JSON descriptor, also uploaded to \textit{software} container, which is then transformed in a user-friendly web interface for users to select the input H\&E WSI.

When the user launches the algorithm initiating the virtual staining, its execution is managed by a job scheduling system based on SLURM, which launches the Singularity image with the corresponding inputs. Upon completion, the generated stains could be directly visualized within the web UI.

Cytomine is optimized for the displaying of multiple instances of aligned WSIs allowing for simultaneous visualization of stains. This functionality significantly enhances the ability to compare and analyze different staining results within a unified interface, providing a powerful tool for digital pathology and related research fields.

\subsection{Experimental configurations}
To ensure reproducibility, it is important to note that all experiments conducted in this study utilized the same architecture for the encoder, decoder, and discriminator. The number of parameters was aligned with those specified in \cite{ComboGAN2017} and implemented using the PyTorch library (version 2.2.0 with CUDA v12.1 and cuDNN v8.902)~\cite{pytorch}. All training sessions were performed using 2048x2048 tiles resized to 512x512 tiles (no overlap) from the Crohn's dataset, as discussed in Section~\ref{method:data}, in either paired or unpaired settings. The models employed an Adam optimizer with parameters $\beta_1 = 0.5$ and $\beta_2 = 0.999$, and a batch size of 6. We used only random flip and random rotation (data augmentation strategies). Each training epoch contains 728 iterations. Each training was conducted on a single NVIDIA A100 80GB GPU. The experimental setup is outlined below.

\subsubsection{Enhanced performance and efficiency in multi-virtual staining using unified H\&E encoder}\label{exp:oneHE}

\begin{figure}[ht]
    \centering
    \includegraphics[width=1\linewidth]{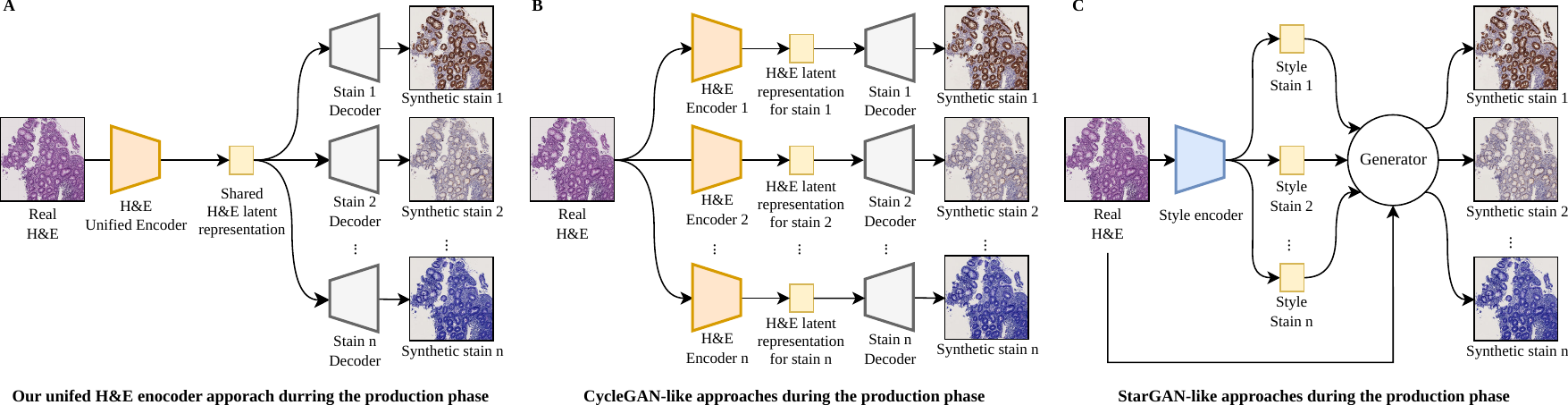}
    \caption{\textbf{Comparison of H\&E staining-based methodologies for virtual stain generation in computational histopathology during production phase.} Panel \textbf{A} illustrates the proposed unified H\&E encoder approach, adapting the ComboGAN~\cite{ComboGAN2017} approach to virtual staining, employing a single encoder and multiple decoders to generate various synthetic stains, thereby optimizing computational efficiency and scalability (to maintain focus on comparative methodology details on XAI capabilities are presented in Figure.\ref{fig:virtual-XAIinfrence}). Panel \textbf{B} depicts the traditional CycleGAN-like methodologies~\cite{goodfellow2014generative,cycleGAN}, which use multiple separate encoders and decoders for each stain, increasing model complexity and computational demand. Panel \textbf{C} showcases the StarGAN-like approaches~\cite{stargan,stargan_2,UMDST,ZHANG2022MVFStain}, using a style encoder and a single generator for multiple stains. While this architecture simplifies the model, it requires substantial computational resources and does not scale effectively, particularly as the number of stains increases (more stains bigger generator), and still necessitates loading the large generator even for a subset of stains, leading to inefficiencies. The unified H\&E approach in panel \textbf{A} represents a significant advancement by reducing the need for multiple models and facilitating quicker, more resource-efficient processing. This model is able to produce only the required stains, loading minimal model components into memory, which reduces hardware requirements and computational costs in cloud-based deployments.}\label{fig:unifed_vs_cyclegan_vs_stargan}
\end{figure}

In Table~\ref{table:oneHE}, we trained two different approaches—our unified method and CycleGAN (refer to Figure~\ref{fig:unifed_vs_cyclegan_vs_stargan})—. For CycleGAN, a separate model was trained for H\&E to each of the different stains, with a total of eight stains. This involved 16 encoders, 16 decoders, and 16 discriminators. Each model underwent 75 epochs at a fixed learning rate of $2 \times 10^{-4}$, followed by 75 decay epochs with a linearly reducing rate, totaling 150 epochs per stain (1200 epochs overall). The loss weights were set to $\lambda_{\text{cyc}}=10$ and $\lambda_{\text{adv}}=1$, with no regularization as $\alpha=0$ and $\beta=0$. 

In contrast, our approach involves simultaneous training for H\&E to the eight different stains, using a total of 9 encoders, 9 decoders, and 9 discriminators. The training consists of 500 epochs at a fixed learning rate of $2 \times 10^{-4}$, followed by 500 decay epochs with a linearly reducing rate. The loss weights and regularization settings are identical to those used in the CycleGAN models.

The values presented in Table~\ref{table:oneHE} represent the mean tile-wise (no overlap) MSE for each stain tested on the Crohn's dataset (refer to Section~\ref{method:data}). These MSE values are computed for both approaches -- our unified method and CycleGAN --, being reported in Table~\ref{table:oneHE}.

\subsubsection{Impact of incorporating IHC loss functions and H\&E regularization on stain synthesis quality}\label{exp:immunoLoss}

In Table~\ref{table:immunoLoss}, the training involved 9 encoders, 9 decoders, and 9 discriminators. The model underwent 500 epochs at a fixed learning rate of $2 \times 10^{-4}$, followed by 500 decay epochs with a linearly reducing rate, summing up to a total of 1000 epochs (paired and unpaired). The impact of incorporating different loss functions and regularization was studied, specifically:

\begin{itemize}
    \item \(\mathcal{L}_{\text{H\&E}}\) (\checkmark): This regularization was applied at the end of each iteration, where the cycle consistency losses \(\mathcal{L}_{\text{cyc},i}\) from the 8 components of the Crohn dataset were summed and averaged. The loss weights were set as $\lambda_{\text{cyc}}=10$ and $\lambda_{\text{adv}}=1$, with $\alpha=0$ and $\beta=0$.
    
    \item \(\mathcal{L}_{\text{IHC}}\) (\checkmark): For the IHC-specific loss, $\lambda_{\text{cyc}}=10$ and $\lambda_{\text{adv}}=1$ were maintained, and values of $\alpha$ and $\beta$ were computed as detailed in Section~\ref{method:Lcyc_Lihc}.
    
    \item Combined \(\mathcal{L}_{\text{H\&E}}\) (\checkmark) and \(\mathcal{L}_{\text{IHC}}\) (\checkmark): Both H\&E regularization and IHC loss were applied similarly as described above, with $\alpha$ and $\beta$ values computed according to the method outlined in Section~\ref{method:Lcyc_Lihc}.
\end{itemize}

The values presented in Table~\ref{table:immunoLoss} represent the MSE, PSNR abd SSIM computed at the WSI level. These metrics are calculated for WSIs reconstructed with $0\%$ overlap and represent the mean values across all eight different stains of the Crohn dataset. Further details on the validation protocol are provided in Section~\ref{sup:quant_validation}.

\subsubsection{Comparison of our model's performance across different magnifications}\label{exp:mag}

In Table~\ref{table:mag}, we evaluated the performance under both paired and unpaired settings using specific magnifications. This involved 9 encoders, 9 decoders, and 9 discriminators. The model underwent 500 epochs at a fixed learning rate of $2 \times 10^{-4}$, followed by 500 decay epochs with a linearly reducing rate, summing up to a total of 1000 epochs (paired and unpaired). The magnifications tested were:
\begin{itemize}
    \item x10 with an original tile size of 2048x2048 pixels, which corresponds to approximately $450.56 \times 450.56 \mu m$,
    \item x20 with an original tile size of 1024x1024 pixels, approximately $225.28 \times 225.28 \mu m$,
    \item x40 with an original tile size of 512x512 pixels, approximately $112.64 \times 112.64 \mu m$.
\end{itemize}

All images are resized to 512x512 for training, following the configuration detailed in Section~\ref{exp:immunoLoss}. This configuration employs combined loss functions $\mathcal{L}_{\text{H\&E}}$ and $\mathcal{L}_{\text{IHC}}$, with parameters $\lambda_{\text{cyc}}=10$ and $\lambda_{\text{adv}}=1$. The performance metrics, listed in Table~\ref{table:mag}, include MSE, PSNR, SSIM. These metrics are computed on WSIs reconstructed with $0\%$ overlap (mean values across all eight different stains of the Crohn dataset). It is important to note that training at a magnification of x40 and testing at x10 requires resizing the synthetic x40 WSI to match the size of the x10 slide. After resizing, metrics are calculated to compare the ground truth slide at x10 with the resized slide. This procedure is applicable to other magnifications as well. Further details on the validation protocol between two WSIs are provided in Section~\ref{sup:quant_validation}.

\subsubsection{Effects of various regularization techniques on unpaired virtual staining performance}\label{exp:compReg}

In Table~\ref{table:compReg}, we evaluated the performance under an unpaired setting using x10 magnification (original tile size of 2048x2048 pixels, corresponding to approximately $450.56 \times 450.56 \mu m$), resized to 512x512. The first row details our approach using the configuration described in Section~\ref{exp:oneHE}. The second row uses the same configuration, combining IHC loss functions with H\&E regularization, as referenced in Section~\ref{exp:immunoLoss}. For subsequent rows, whenever a specific stain regularization is applied, the parameters $\lambda_{\text{cyc}}=10$, $\lambda_{\text{adv}}=1$, and values for $\alpha$ and $\beta$ are computed according to the method outlined in Section~\ref{method:Lcyc_Lihc}. Additionally, $\mathcal{L}_{\text{idt}}=1$, $\mathcal{L}_{\text{lat}}=1$, or $\mathcal{L}_{\text{fwd}}=1$ may be applied, with detailed descriptions of each stain regularization found in Section~\ref{method:stain_regs}.


\section*{Acknowledgments}
This work was granted access to the HPC resources of IDRIS through the allocation number 2023-AD011014513, provided by GENCI. We extend our gratitude to Mr. Blain Pascal, a technician in the Pathological Anatomy Department at Robert Debré Hospital, for his invaluable assistance with the retrieval of the whole slide images. Additionally, we thank Mrs. Dalal Yahiaoune, a clinical research assistant at Robert Debré Hospital, for her help in retrieving, anonymizing, and organizing the data, as well as for her technical support. We are also grateful to the French network for rare digestive disorders for funding the digitization of the slides.

\section*{Author contributions statement}
OM led the development, prototyping, and validation of the pipeline modules, incorporating both traditional and interpretable deep learning techniques using Python. OM also crafted both the qualitative and quantitative analyses by generating the figures and illustrations for this article. SI significantly contributed to validating the XAI methodology, as well as deploying cloud-based computational histopathology and dockerizing the pipeline. RD coordinated, supervised and validated the multi-stain generation methodology and the overall image processing pipeline, further enhancing the design and validation of the pipeline’s performance and its explainable features to ensure they were effective and interpretable for users. BD handled the preparation, inspection, and scanning of the Crohn’s disease multi-stain dataset, providing continuous feedback to improve the methodology. Collectively, OM, SI, RD, BD, HJ, and MC worked together on the study’s design in the context of Crohn’s disease and the interpretation and the reporting of the results. 

\section*{Data and code availability statement}
Upon acceptance of the paper, the virtual staining dataset and code for experiments reproducibility will be accessible through a permanent public link. 

\section*{Competing interests}
The authors declare no competing interests.

\bibliographystyle{unsrt}  
\bibliography{references}  

\newpage
\section{Supplementary materials}\label{sec:extended}

\subsection{Generalization to different stains types}

\begin{figure}[H]
    \centering
    \includegraphics[width=1\linewidth]{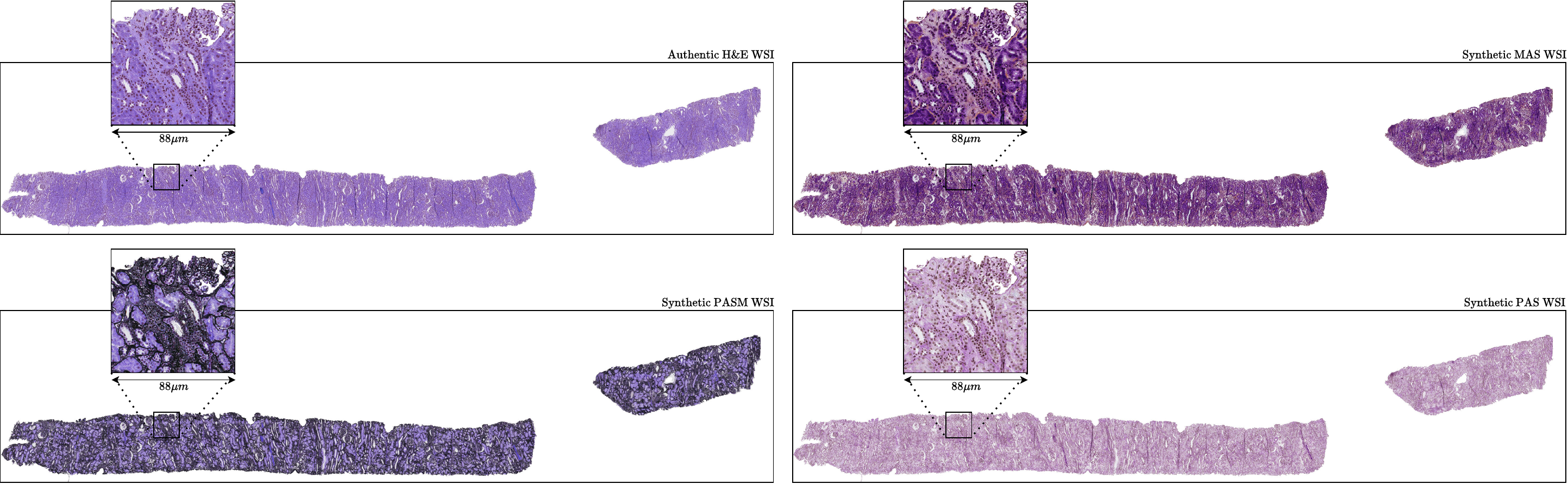}
    \caption{\textbf{Multi-virtual staining results on kidney slide N\textdegree5 from AHNIR dataset.} Showing the high quality synthetic stains generated using our method.}
    \label{fig:demo_kidney}
\end{figure}

The ANHIR~\cite{AHNIR} dataset includes five sets of high-resolution human kidney tissue slides, each containing four slides of consecutive tissues stained with different types (H\&E, MAS, PAS, and PASM staining). Although these slides are structurally similar, they are not pixel-level paired and all are magnified at x40.

In accordance with the experimental setup of UMDST~\cite{UMDST}, four sets (kidney 1, kidney 2, kidney 3, kidney 4) were used as the training sets, and the fifth set (kidney 5) was reserved for testing. From the UMDST protocol, the H\&E stained slide from kidney 1 was omitted due to its distinct color variation compared to the other sets. For processing, slides were tiled into 256x256 images with an overlap of 192 pixels.

Our method entailed simultaneous training across the three different stains from H\&E using four encoders, four decoders, and four discriminators. The training involved 150000 iterations at a fixed learning rate of \(2 \times 10^{-4}\), followed by another 150000 iterations with a linearly reducing rate, totaling 300000 iterations. An Adam optimizer with parameters \(\beta_1 = 0.5\) and \(\beta_2 = 0.999\), and a batch size of 1 was employed (matching UMDST~\cite{UMDST}). Only random flip and random rotation strategies were used for data augmentation. This training was performed on a single NVIDIA A100 80GB GPU. The loss weights were set at \(\lambda_{\text{cyc}}=10\) and \(\lambda_{\text{adv}}=1\), with \(\alpha=0\), \(\beta=0\), and \(\mathcal{L}_{\text{H\&E}}\) applied every three iterations. Here, the cycle consistency losses \(\mathcal{L}_{\text{cyc},i}\) from the components of the kidney dataset were summed and averaged. \(\mathcal{L}_{\text{idt}}=0\), \(\mathcal{L}_{\text{lat}}=0\), and \(\mathcal{L}_{\text{fwd}}=0\) were also included in the model.

Testing on kidney 5 was performed as shown in Figure \ref{fig:demo_kidney} with qualitative results presented in Table \ref{table:kidney}. To benchmark against the state-of-the-art, the Contrast Structure Similarity (CSS) metric\cite{UMDST,CSS1,CSS2} was employed, as reported in Table \ref{table:kidney}.

\begin{table}[H]
\centering
\caption{\textbf{Comparative analysis of CSS for different staining methods across models.} This table presents the CSS metrics for various computational methods when applied to human kidney tissue slides stained with H\&E, MAS, PAS, and PASM. Each method's performance is evaluated in terms of overall CSS, tile output, WSI-compliant output and evaluation, XAI capabilities, and scalability. The results highlight our method's superior ability to address the challenges of multi-virtual staining, with higher CSS values signifying enhanced preservation of structural similarity across different stains.}
\label{table:kidney}
\begin{tabular}{ccccccccc}
\toprule
\textbf{Method} & \textbf{MAS} & \textbf{PAS} & \textbf{PASM} & \textbf{Overall}                                                                   & \textbf{Tile-output} & \textbf{WSI-compliant} & \textbf{XAI} & \textbf{Scalability} \\ 
\midrule
MUNIT~\cite{MUNIT}            & 0.145& 0.115& 0.110& 0.123$\pm$0.015                                              & \checkmark &$\times$ & $\times$ &  $\times$\\
FUNIT~\cite{FUNIT}            & 0.332& 0.318& 0.246& 0.298$\pm$0.037                                              & \checkmark &$\times$ & $\times$ &  $\times$\\
StarGAN~\cite{stargan,stargan_2}         & 0.520& 0.543& 0.491& 0.518$\pm$0.021                                   & \checkmark &$\times$ & $\times$ &  $\times$\\
UGATIT~\cite{UGATIT}          & 0.584& 0.510& 0.399& 0.497$\pm$0.076                                              & \checkmark &$\times$ & $\times$ &  $\times$\\ 
UMDST~\cite{UMDST}            & \underline{0.682} & \underline{0.674} & \textbf{0.600} & \underline{0.652$\pm$0.036} & \checkmark &$\times$ & $\times$ &  $\times$\\ 
\midrule
Ours            & \textbf{0.797} & \textbf{0.784} & \underline{0.536} & \textbf{0.705$\pm$0.120}                     & \checkmark &\checkmark & \checkmark& \checkmark \\ 
\bottomrule
\end{tabular}
\end{table}

Table \ref{table:kidney} highlighting superior performance and the generalization capabilities when compared to state-of-the-art methodologies, particularly in MAS, PAS, and PASM staining. Although the performance in PASM staining is not as high, this is expected given the inherent challenges in accurately keep all the H\&E morphological features (since  CSS metric is computed between the H\&E and the PASM stain). PASM staining tends to obscure some morphological features due to its use of black coloration, thus, our model learn to do so too (it is a feature not a bug). But if the intention is to preserve these features (not realistic PASM but with all H\&E features) a forward loss strategy can be used, as described in UMDST \cite{UMDST}. However, this introduces a trade-off between maintaining morphological detail and achieving accurate staining. Furthermore, our approach is scalable during training and inference and uniquely integrates XAI capabilities, not addressed by other methodologies, which are critical in a clinical context.

\newpage
\subsection{Validation protocol for virtual staining}\label{sup:quant_validation}
\subsubsection{Quantitative evaluation}

To address the inherent limitations of patch-level evaluation in virtual staining, such as restricted contextual information and potential inconsistencies across different tissue regions, we developed an adapted validation protocol. Traditional metrics often fail to capture the nuanced discrepancies that can occur across various regions of a tissue slide, leading to an incomplete assessment of staining quality. Our protocol, by contrast, incorporates both PSNR and SSIM to comprehensively assess the quality of WSIs. These metrics are crucial for evaluating the fidelity and structural integrity of virtually stained images. Furthermore, MSE metric is specifically employed to provide a quantitative assessment at the tissue pixel-level, significantly enhancing the precision in evaluating staining accuracy.

The use of a paired dataset, where each virtual stain is directly compared to a chemically stained ground truth counterpart (GT stain WSI), is pivotal. This pairing ensures that each evaluation metric not only measures the error or similarity in isolation but does so in a context that reflects true biological and clinical scenarios, ensuring the relevance and applicability of the findings.

\begin{figure}[H]
    \centering
    \includegraphics[width=0.9\linewidth]{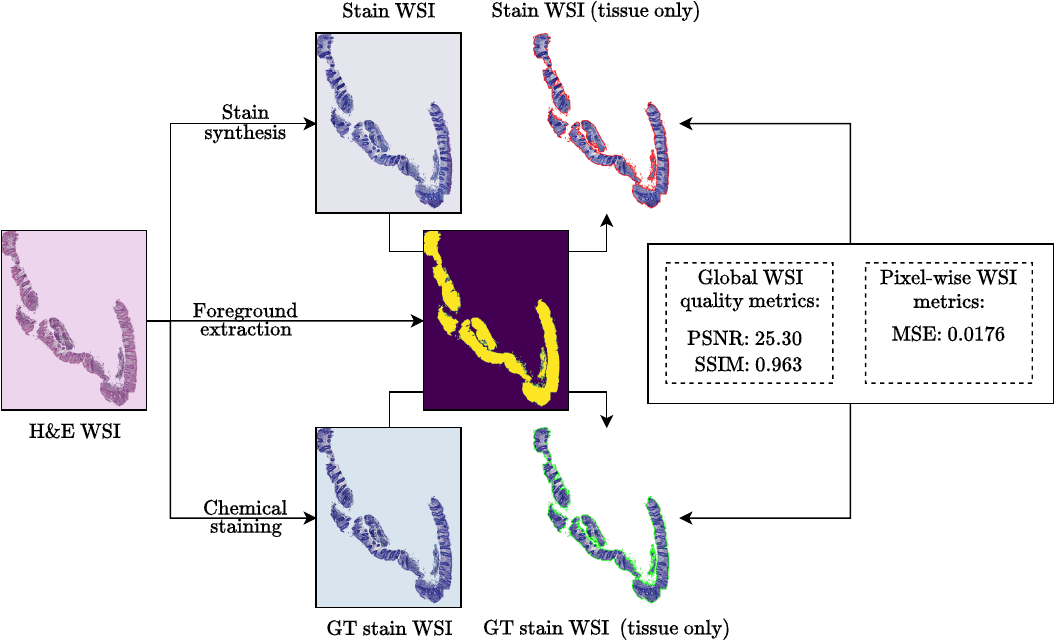}
    \caption{\textbf{Evaluation protocol for virtual staining performance.} Workflow diagram illustrating the validation process for virtual staining techniques. The process begins with an H\&E stained whole slide image (H\&E WSI), from which the foreground is extracted. This image undergoes virtual staining to produce the Stain WSI, which is then compared to the chemically stained ground truth WSI (GT stain WSI). The evaluation metrics include PSNR and SSIM for assessing overall image quality, and MSE for pixel-wise accuracy, indicating the effectiveness of the staining simulation.}
    \label{fig:eval_protocol}
\end{figure}

The refined validation protocol involves several steps. Initially, an H\&E stained WSI is processed to extract the foreground, effectively distinguishing the tissue from the background. Subsequent virtual staining algorithms synthesize the stain, producing a stain WSI that is then compared against the ground truth stain WSI obtained from chemical staining. This comparison is essential for assessing the virtual staining's performance across entire slides and at the pixel level as illustrated in Figure.\ref{fig:eval_protocol}. Through these metrics, our protocol addresses critical gaps in existing evaluation methods and sets a clear validation of virtual staining technologies in pathology.

\subsubsection{Qualitative evaluation}

In our study, we recognize the importance of qualitative evaluation alongside quantitative metrics, particularly from a pathological perspective. Despite utilizing a paired dataset, qualitative assessment remains crucial for verifying the applicability and accuracy of our virtual staining techniques from a clinical standpoint.

\begin{figure}
    \centering
    \includegraphics[width=\linewidth]{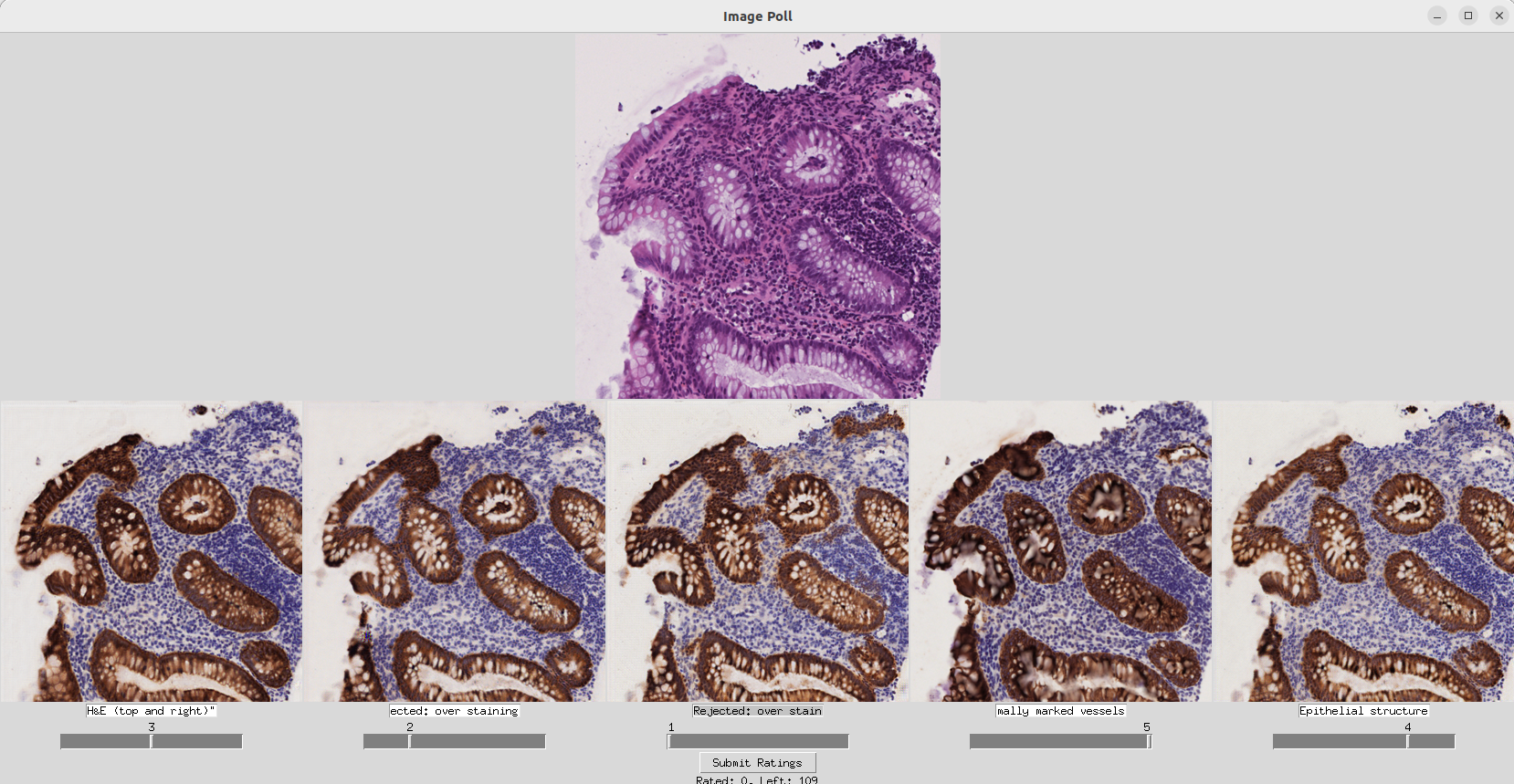}
    \caption{\textbf{Software for poll results and feedback collection of pathologist ratings on staining quality.} We show the original H\&E image at the top, followed by a sets of virtual stains in different conditions including the ground truth randomly showed. Pathologist was asked to rate each image based on the clarity and preservation of morphological details 1 "worst" 5 "best" with a feedback.}
    \label{fig:poll}
\end{figure}

In Figure.\ref{fig:poll}, we conducted a poll involving 26 images stained with AE1AE3, where a pathologist was shown the original H\&E image alongside virtual staining results. These included images processed through real chemical staining (ground truth) and those generated via our paired and unpaired DL models. Pathologist was instructed to rate the images on a scale from 1 (worst) to 5 (best) and provide feedback.

The outcomes of our study were somewhat counter-intuitive. In the assessment of 26 AE1AE3-stained images, the ground truth images, which involved actual chemical staining, generally scored lower than those from both the paired and unpaired settings. Specifically, the ground truth images received an average score of $2.69 \pm 1.46$. In contrast, images from the paired setting, where virtual staining was trained on paired data, scored slightly higher at $3.11 \pm 1.63$. Most notably, the unpaired setting, involving virtual staining trained without paired data, performed the best with an average score of $3.42 \pm 1.65$. This suggests an unexpected performance trend where the virtually generated stains were preferred over the actual chemical stains, indicating a discrepancy in quality perception between the traditional and computational methods.

\begin{figure}
    \centering
    \includegraphics[width=1\linewidth]{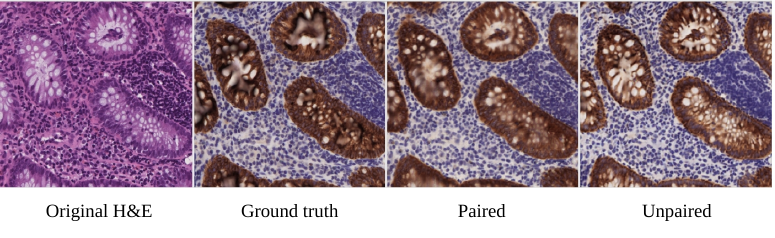}
    \caption{\textbf{Morphological detail comparison in H\&E stained images.} This figure shows a closer view of the morphological features in the original H\&E stain (left) versus the ground truth, paired, and unpaired virtual stains. The comparison highlights the impact of water-like blur in chemical stains and its reduction in virtual stains, aiding in the qualitative assessment by pathologists.}
    \label{fig:morph}
\end{figure}

Upon analyzing the pathologists' feedback, a critical observation was made, as illustrated in Figure.\ref{fig:morph}. It appears that a water-like blur inherent in the chemical staining process tended to obscure the morphological details of the tissue. This issue was less pronounced in the images from the paired and unpaired settings.

Notably, the unpaired model displayed superior preservation of morphological features. This is likely because, during training, the model does not directly correlate the H\&E images with specific stains, allowing it to learn where to place stains effectively without replicating the blurring seen in the ground truth. Conversely, the paired model, learning from the blurred ground truth images, tends to reproduce similar artifacts, thus inheriting and replicating these biases. These findings underscore that unpaired training can provide more generalized and unbiased results. While objective metrics might suggest lower performance compared to the paired settings, the qualitative benefits from a pathological perspective are worth noting. The unpaired setting yields visual results that surpass the ground truth (when there is a blur effect), offering enhanced clarity and detail that are crucial for accurate medical diagnosis.

\subsection{Testing IDs for crohn's dataset}\label{sup:crohn_data}

For reproducibility, the specific slide IDs assigned to the test set for the dataset are designated as follows: AE1AE3 test slides include numbers 9, 13, 19, 21, 22, 24, 26. For CD3, the test slides are 0\_0, 0\_1, 0\_2, 5\_0, 5\_1, 11. CD8 test slides are numbered 0, 1, 12, 14, 15, 17. CD15 slides identified for testing are 6\_0, 6\_1, 6\_2, 8\_0, 8\_1, 8\_2, 11\_0, 11\_1, 11\_2, 14\_0, 14\_1, 14\_2. CD117 includes test slides 2\_0, 2\_1, 2\_2, 8\_0, 8\_1, 8\_2, 10\_0, 10\_1, 10\_2, 13\_0, 13\_1, 13\_2. CD163 comprises slides 6, 9, 12, 15, 25, 27. D2-40 test slides include 3\_0, 3\_1, 3\_2, 3\_3, 7, 9. Lastly, GIEMSA slides for testing are numbered 3\_0, 3\_1, 3\_2, 8\_0, 8\_1, 8\_2, 9\_0, 9\_1, 9\_2, 9\_3, 11\_0, 11\_1.

\end{document}